\documentclass[acmtog,authorversion,nonacm]{acmart}

\pdfoutput=1

\usepackage{multirow}
\newcommand{\eg}{{\textit{e.g., }}}
\newcommand{\ie}{{\textit{i.e., }}}

\begin{document}
\title{Procedural Urban Forestry}

\author{Till Niese}
\affiliation{%
\institution{University of Konstanz}
}

\author{S\"oren Pirk}
\affiliation{%
\institution{Google Brain}
}

\author{Matthias Albrecht}
\affiliation{%
\institution{University of Konstanz}
}

\author{Bedrich Benes}
\affiliation{%
\institution{Purdue University}
}

\author{Oliver Deussen}
\affiliation{%
\institution{University of Konstanz}
}

\renewcommand\shortauthors{Niese, T. et al.}

\begin{abstract}
The placement of vegetation plays a central role in the realism of virtual scenes. We introduce procedural placement models (PPMs) for vegetation in urban layouts. PPMs are environmentally sensitive to city geometry and allow identifying plausible plant positions based on structural and functional zones in an urban layout. PPMs can either be directly used by defining their parameters or can be learned from satellite images and land register data. Together with approaches for generating buildings and trees, this allows us to populate urban landscapes with complex 3D vegetation. The effectiveness of our framework is shown through examples of large-scale city scenes and close-ups of individually grown tree models; we also validate it by a perceptual user study.
\end{abstract}

%
%

\ccsdesc[500]{Computing methodologies~Shape analysis}
\ccsdesc[300]{Theory of computation~Grammars and context-free languages}
\ccsdesc[100]{Theory of computation~Rewrite systems}
%
%

\keywords{Urban Models, Vegetation, Procedural Generation, Urban Forestry}

\begin{teaserfigure}
  \centering
   \includegraphics[width=\linewidth]{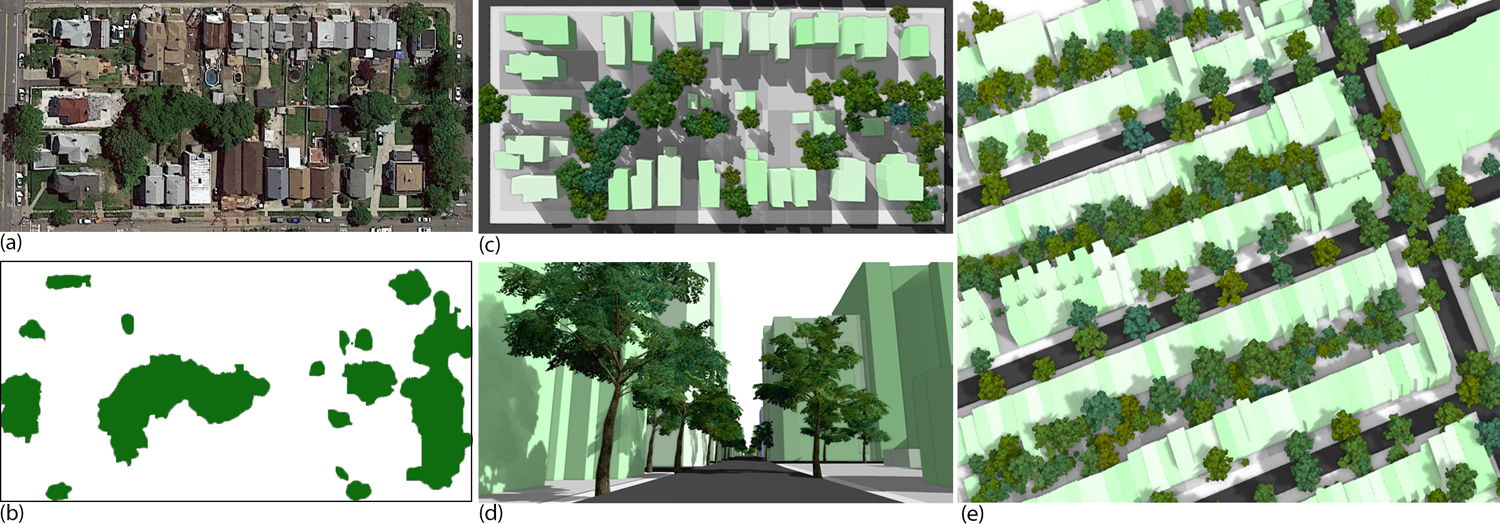}
   \vspace{-5mm}
  \caption{Steps of our learning-based plant population method: we use satellite images (a) and predict coverage maps for vegetation (b). 
  We use these maps to identify plant regions for reconstruction (c) and to learn parameters for our procedural models when populating new virtual cities with complex plants (d), which significantly increases the realism of urban landscapes (e).}\label{fig:teaser}
  \vspace{0mm}
\end{teaserfigure}

\maketitle

\section{Introduction} \label{sec:intro}
The visual simulation of urban models and generation of their 3D geometries are important open problems in computer graphics that have been addressed by many approaches. Existing methods range from fa\c{c}ades, buildings, city block subdivisions, to entire cities with viable street and road systems. 
Synthetically generated city models already exhibit a high degree of realism. However, cities are immersed in vegetation, but only very little attention was dedicated to the interplay of urban models and vegetation in computer graphics. Ecosystem simulations have been considered by many approaches. The prevailing algorithms use plant competition for resources as the main driving factor of their evolution either on the level of entire plants~\cite{Deussen98} or on the level of branches~\cite{Makowski:2019:SSM:3306346.3323039}. Unfortunately, these approaches fail in urban areas, because urban trees have only limited space available to compete for resources, and they are heavily affected by surrounding urban areas as well as by human intervention. 

The term urban forest refers to vegetation in urban areas~\cite{miller2015urban}. Vegetation has many practical functions: it limits and controls air movement, solar radiation, and heat, humidity, and precipitation, it can also block snow and diminish noise. Moreover, an important function of vegetation is to increase city aesthetics. Urban forests are not planted at once but managed over time. Dead trees are removed and new trees are planted. Living trees are pruned for visibility or utility services. In contrast to real cities, we face a different situation in computer graphics. An existing algorithm generates a city model without vegetation and we need to find suitable locations for individual trees. Simulating urban forest evolution, \ie by using the algorithm by Benes et al.~\shortcite{10.1145/1944745.1944773}, is time-consuming and difficult to control.

In this paper, we introduce a procedural method for the advanced placement of vegetation to increases the overall realism of urban models. We are inspired by urban rules that control which trees and bushes can be planted where and how tall they are allowed to grow. These rules vary for individual areas; they are relaxed in industrial zones, people also have more flexibility in their properties, but they are enforced in public zones of a city and around important landmarks. We, therefore, introduce procedural placement models -- strategies for generating plant positions --  along with parameters to enable an automatic placement of vegetation, faithful to the characteristic features of plant distributions within the different municipality zones of a city. We show that placement models and parameters together provide an efficient means of controlling the interactive modeling of urban landscapes.

Moreover, we can populate city models not just with static tree geometry, but with dynamic models of plants that can grow and change their shape in response to environmental changes or human intervention. This allows us to apply simulation models that describe how a city, or its areas would change if more or less effort could be spent on maintaining them. Having such dynamic urban ecosystems allows users to visually predict and control the effects of gardening in a city and also helps to make such models more realistic since they inhibit decay and different levels of order.

While procedural placements can be used directly to populate urban layouts, we also show that placement models can be used to learn plant distributions of real cities. We use satellite images and land register data to train deep neural networks to learn the distributions of trees and other plants in the parameter space of our procedural placement models. While placement models act as a strong prior to regularize finding plausible placements, learning parameter values also enable users to efficiently author scenes through intuitive parameters. The example in Fig.~\ref{fig:teaser} shows a satellite image (a) and the predicted coverage map (b). We use coverage maps to identify areas where to place vegetation (c) and to learn parameters of our procedural models. Once the parameters are obtained we can automatically populate city models with complex models of plants (d) to increase their realism (e).

Our main contributions are: (1) we advance the state-of-the-art in modeling vegetation in urban landscapes by introducing a procedural modeling framework that is based on the idea to factorize the complexity of plant placement into manageable components; (2)~we introduce a set of procedural placement models along with their parameterization to capture a large variety of placement patterns; (3)~we use a novel pipeline for learning plant distributions in cities from satellite data; we convert satellite images into coverage maps and then learn the placement parameters of our procedural models. 

\section{Related Work} \label{sec:rw}
Only recently researchers started exploring approaches to model virtual environments with realistic traits of real urban landscapes~\cite{Smelik:2014:SPM:3071829.3071833}. Here, we focus on the involved aspects of plant and urban modeling, ecosystems as well as learning-based methods.

\textbf{Urban Modeling:} urban structures are often modeled proceduraly~\cite{4497893}. In their seminal paper, Parish and M\"uller~\shortcite{Parish2001PMC} used L-systems to model complex cities and Wonka et al.~\shortcite{Wonka:2003:IA:882262.882324} applied split grammars to procedurally define buildings that were later extended by using subdivision~\cite{Mueller:06:SIGG} and by more advanced operations~\cite{Schwarz:2015:CGAPP}. Purely procedural models of infinite cities were introduced by Merrell and Manocha~\shortcite{Merrell:11,Merrell:08}, the procedural modeling of street layouts has been described by using vector fields ~\cite{10.1145/1399504.1360702}. Similarly, procedural approaches have been successfully applied to modeling fa\c{c}ades~\cite{Mueller:07:SIGG}. Urban modeling has been combined with urban simulation to generate viable cities ~\cite{Vanegas09,Vanegas2010}, and city growth~\cite{Weber09}. 

\textbf{Inverse Procedural Modeling:} our approach is related to inverse procedural models in that it learns plant placement from real cities and attempts to transfer it to synthetic ones by fitting parameters of a procedural model. An inverse procedural model for fa\c{c}ades has been introduced by AlHalawani et al.~\shortcite{Al:13:EG} and Wu et al.~\shortcite{Wu:14:Sigg}, variations from a procedurally encoded single layout can be generated by the work of Bao et al.~\shortcite{Bao:13:PFV}, 
the layered nature of fa\c{c}ades has been used for inverse procedural modeling in~\cite{Li:11:ICCV,Ilcik:2015:LPD:2816723.2816742}, exploiting structural symmetries was done in~\cite{Dang2014}, and interactive alterations of shape grammars were utilized in~\cite{Dang:2015:IDP:2816795.2818069}. Buildings can be encoded as L-systems by using the inverse procedural approach from~\cite{Vanegas:10:CVPR}, modeled by using a procedural connection of structures~\cite{Bokeloh2010connection}, or through binary integer programs~\cite{10.1145/3130800.3130823}. Finding the parameters of procedural models from existing data was investigated by Talton et al.~\shortcite{Talton2011metropolis} who used expressions of L-system strings of modules to fit a generated structure to an input. Ritchie et al.~\shortcite{Ritchie2015controlling} attempt to control procedural programs and procedural models using stochastic Monte Carlo methods.
Structural patterns can be encoded by using the approach of Yeh et al.~\shortcite{Yeh:2013:STP:2421636.2421639} or encoded as L-systems by the work of \v{S}\v{t}ava et al.~\shortcite{Stava2010}.
Recently, trained deep neural networks have been combined with inverse procedural modeling to allow for the interactive design of buildings by using sketches~\cite{nishida2016interactive}. Inverse procedural modeling has also been used to generate entire urban layouts in~\cite{Vanegas:2012,Martinovic2013bayesian}.

\textbf{Plant Modeling:} research has long focused on defining plausible branching structures based on fractals~\cite{Aono:CGaA:84, Oppenheimer:86} or L-Systems~\cite{Lindenmayer:68, Pwp:GI86}. Other methods focus on rule-based modeling~\cite{Lintermann:99:CGaA}, inverse procedural modeling of trees~\cite{Stava2010,Stava2014}, and finding L-system for branching structures~\cite{10.1145/3394105}. Moreover, sketch-based modeling techniques allow artists to produce plant models interactively and in more nuanced ways~\cite{OkabeSketchTree06,wither:hal-00366289,Ijiri:EG:06}. 
Alternative approaches attempt to reconstruct plant models automatically either from  images~\cite{Tan:2007:ITM:1276377.1276486,Tan:2008:SIT:1409060.1409061}, videos~\cite{Li:2011:MGM:2070781.2024161}, or scanned 3D point clouds~\cite{Tree2016,Livny:2011:Sigg}. Only just recently, several approaches also focus on the dynamic and realistic behavior of plant models, including  growth~\cite{Pirk:2012:CAM:2366145.2366188,Longay:SBIM:12}, the interaction with wind or fire~\cite{Pirk:2017:IWC:3130800.3130814,Pirk:2014:WTC:2661229.2661252}, or as established through realistic materials~\cite{Wang:2017:BMB}. 

Modeling the plants' response to its environment is of utmost importance to obtain realistic branching structures when positioned in groups or alongside obstacles~\cite{Mech:96:SIGG}. Approaches exist to model this phenomenon by considering the self-organization of plants~\cite{Runions07,Palubicki:2009:STM}, through explicitly modeling the plasticity of branches~\cite{Pirk:2012:PTI:2185520.2185546} or through the dynamic adaptation to support structures, as can be observed for climbing plants~\cite{Benes:02:CA,Haedrich:2017}. The growth, decay, and pruning of buds and branches play an eminent role in plant development~\cite{deReffye:1988:PMF:378456.378505}; a phenomenon that is often parameterized in procedural models to develop convincing branching structures~\cite{Stava2014}. 

\textbf{Ecosystems:} various works focus on ecosystem simulation. The seminal paper of Deussen et al.~\shortcite{Deussen98} introduced a competition for resources on the plant level, this approach has been recently extended towards the competition of individual trees in layered ecosystems~\cite{Makowski:2019:SSM:3306346.3323039}. Various approaches attempt to simulate ecosystems considering different phenomena, such as erosion~\cite{cordonnier2017authoring} or even by locally learning plant distributions and using them as interactive brushes~\cite{gain2017ecobrush,10.1145/2766975}. Close to our approach is the work of Benes et al.~\shortcite{10.1145/1944745.1944773} that models urban ecosystems by combining wild ecosystem growth from~\cite{Deussen98} with controlled plant management. However, contrary to our work, the initial plant placement is purely ad hoc and their approach does not allow for procedural plant placement that could be connected with real cities.

\textbf{Learning-based Approaches:} some works have started to explore the capabilities of learning-based methods for scene generation and object placement. While neural networks have shown paramount performance on image classification, synthesis~\cite{8195348, kahn2019}, or inverse texture modeling~\cite{10.1145/3355089.3356516} tasks, properly placing objects into meaningful configurations is still a challenging problem. For arranging scenes, methods need to coherently generate plausible and continuous poses (translation and orientation) of objects and to one another. However, most neural network architectures only allow to operate on fix-sized in- and outputs, which makes placing arbitrary numbers of objects challenging. To this end, Ritchie and Wang~\shortcite{Ritchie2018FastAF}, as well as Wang et al.~\shortcite{10.1145/3306346.3322941}, propose methods for scene generation based on convolutional neural networks, while Zeng et al.~\shortcite{10.1007/978-3-030-01219-9_45} learn to reconstruct buildings by learning parameters of a procedural model.
For outdoor scenes Guerin et al.\shortcite{Guerin:2017:IET:3130800.3130804} and Kelly et al.~\shortcite{KellyGuerreroEtAl_FrankenGAN_SigAsia2018} use generative adversarial networks to author textures for terrain and building details.

While these methods only tangentially relate to our work, they show the capabilities of neural networks for scene generation. Similarly to these methods we combine the advantages of image-based learning techniques with procedural modeling. In particular, we aim at learning the parameters of procedural models with neural networks that allow us to realistically place plants.

\section{Overview} \label{sec:overview}
\begin{figure}[t]
    \centering
    \includegraphics[width=1.0\linewidth]{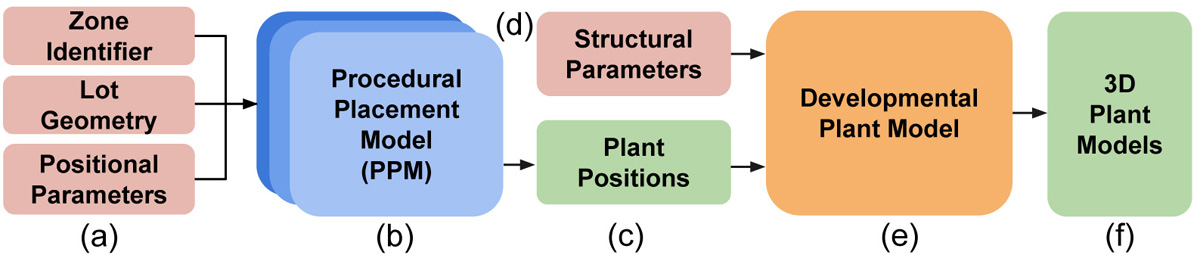}\\
    \vspace{-1mm}
    \caption{
    To place vegetation in urban environments we propose procedural placement models (b) that implement placement strategies for vegetation based on the geometry of individual lots, positional parameters, and a zone identifier (a). After plant positions (c) have been generated we use a developmental model (e) along with structural parameters (d) to jointly grow plants, which results in realistic 3D plant models.}
\vspace{-3mm}
    \label{fig:pipeline}
\end{figure}

Generating plausible vegetation models for virtual urban landscapes faces two major challenges: first, plant placement varies across different functional and demographic zones (Fig.~\ref{fig:pipeline}a)-- an industrial zone may only have a small number of non-managed plants, while residential areas not only have regularly placed trees alongside roads but also in gardens and parks. The planting rules depend on culture, habits, city rules, etc. and thus are difficult to quantify. Second, plant models need to simulate growth and interaction with their environment to generate vegetation with high visual fidelity. 
Moreover, urban trees are often pruned or may lack resources (water or light) which hinder their growth and affect their structure.

To address these challenges, we propose a two-stage procedural modeling pipeline. First (Fig.~\ref{fig:pipeline}), we introduce PPMs (b) to generate plausible plant positions based on placement strategies and known planting rules for vegetation. A PPM can be defined for each functional or demographic zone of a city (\eg residential, commercial, or industrial) and operates on single lots of land (realty). Each PPM has a different set of rules parameterized by structural and positional parameters that allow us to capture the various kinds of planting patterns found in real cities. Second, once the plant positions are generated, we use a state-of-the-art developmental model (Fig.~\ref{fig:pipeline}, e) for growing plants. Given the plant's location and environment the growth process generates unique and realistic branching structures. 

Finally, we have developed a novel learning-based pipeline for populating models of real cities with vegetation. First, we convert satellite images of urban landscapes to \textit{vegetation coverage maps} by using a style-transfer networks (Fig.~\ref{fig:networks}, b). The coverage maps represent areas that are covered with above-ground vegetation. Second, we learn a mapping from the coverage maps to the parameters of our PPMs (Fig.~\ref{fig:pipeline}, d). Given our pipeline and the parameter values obtained from real satellite images, we can generate vegetation similar to what can be observed in the satellite images. \textbf{}

\section{Planting Rules}\label{sec:zones}
Landscapes are defined by road networks and administrative or functional zones~\cite{waddell2007incorporating} and they can be further classified into rural, exurban, suburban, and urban areas~\cite{miller2015urban}. All the involved plants together form the urban forest, an umbrella term referring to trees, shrubs, and bushes found in urban and suburban areas.

The most common way of adding a tree into an urban forest is by replacing a dead tree. It is quite uncommon that large areas are directly populated by vegetation, such a  situation occurs usually only in newly created developments. When a new neighborhood is built, a city would plant regularly spaced trees and bushes parallel to roads and sidewalks by applying municipal tree ordinances~\cite{grey1995urban} (see also~\cite[pg 254]{miller2015urban}). The neighborhood is subdivided into blocks and blocks into individual lots that are left to the owners to plant the vegetation as needed. Typically, the city only defines certain planting rules such as the distance between individual trees should depend on the tree height, or the distance is derived from the soil the tree requires to survive~\cite{endreny2018strategically}. Trees should not obstruct views at intersections, they should have a certain distance from the curb, and sidewalks~\cite{bloniarz1993designing}. Vegetation must not block house entrances for emergency purposes. These functional restrictions are also combined with aesthetic constraints: vegetation should not be planted in the proximity of windows~\cite{miller2015urban}. Most of these rules are combined into a so-called building activity area (or building envelope) that is an extension of the 2D projection of the building by about 600cm perpendicularly from each wall of the building and 150cm from each driveway.

At the lowest level, we create procedural placement models that seek to position all trees at once. Implementing the above-described spacing rules results in vegetation that is regularly placed along roads and sidewalks and that follows a so-called Poisson-disc distribution (minimal distance requirement) among plants and from the building envelope. We further expand the building envelope to consider aesthetic criteria such as planted vegetation should not obstruct views from windows.

At a higher level, we aim to procedurally generate vegetation for the various types of zones. Therefore, we assume that each urban layout, either real or synthetically generated, can be divided into such zones. Specifically, we use a zonal layout commonly used in urban planning~\cite{waddell2007incorporating,waddell2002urbansim} and urban simulations~\cite{Vanegas09,Vanegas2010,Weber09} and divide an urban layout into five zones: 1)~\textit{residential} includes houses and buildings where people live, 2)~\textit{commercial} consists of businesses such as department stores, malls, and small stores, 3)~\textit{industrial} zones include factories and other production services, 4) \textit{street} zones, which describe areas next to roads. Additionally, we add a fifth category (5)~\textit{other} that includes parks, non-managed areas, areas close to railroads, unassigned areas, etc. 

\begin{figure}[t]
    \centering
    \includegraphics[width=\linewidth]{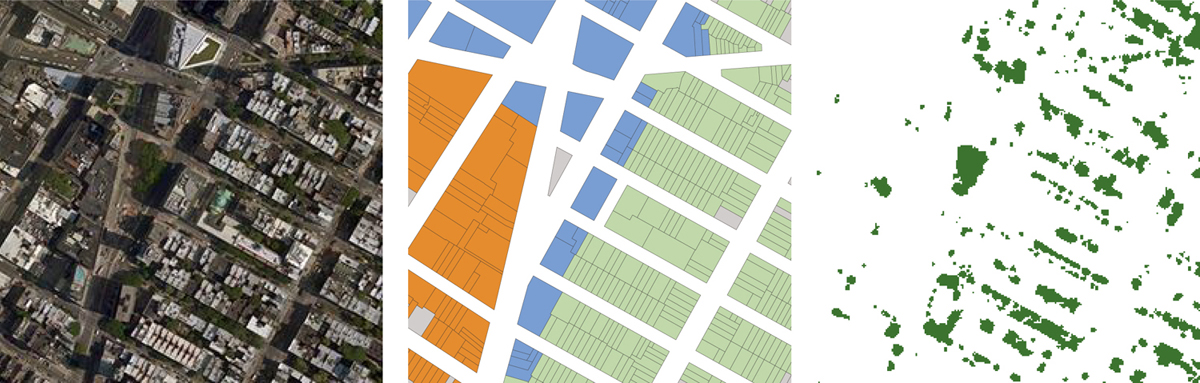}\\
    \vspace{-2mm}
    \caption{Urban layout: satellite images (left), zone data for individual lots (middle), and coverage maps (right) are available in public datasets. We use zone data and lot geometry as inputs to our procedural models and learn to predict their parameter values from the coverage maps.}
    \label{fig:zones}
\vspace{2mm}
    \centering
    \includegraphics[width=\linewidth]{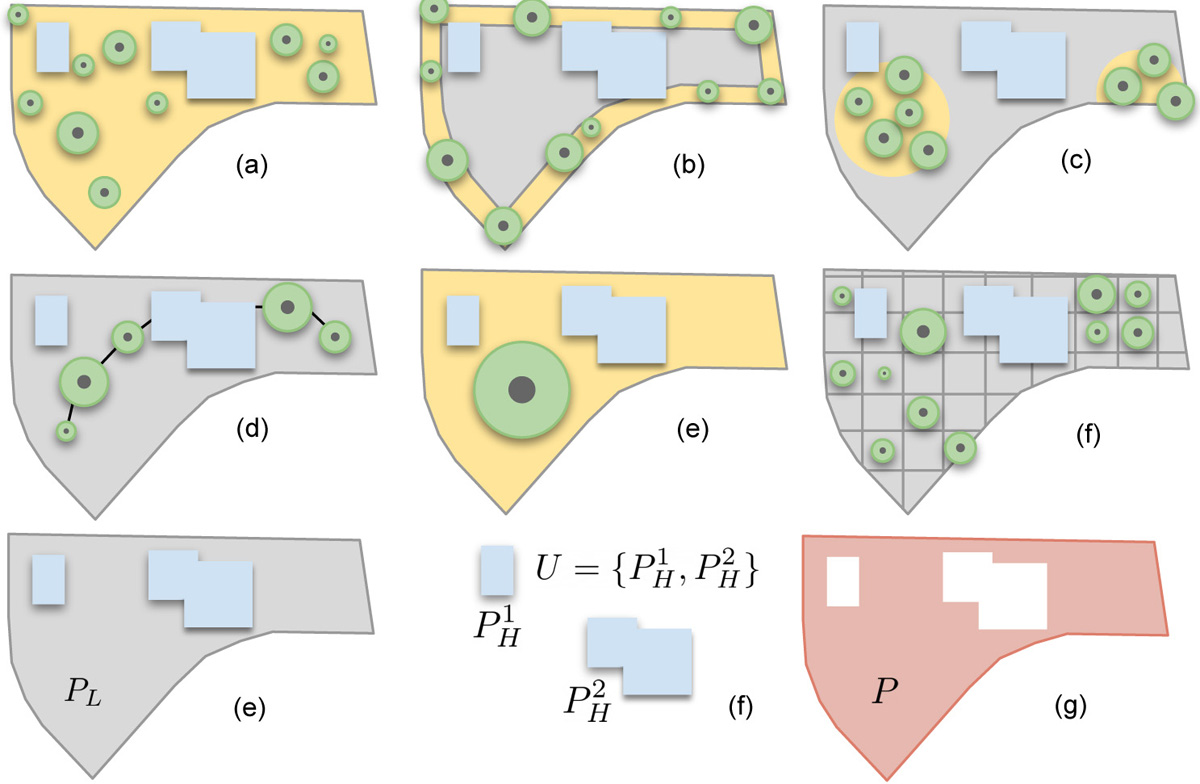}\\
    \vspace{-3mm}
    \caption{Placement strategies illustrated for a single lot: random~(a), boundary~(b), cluster~(c), equidistant~(d), single~(e), regular~(f). While the PPM only places plant positions within a lot, the resulting plants can grow over its boundaries. For the strategies random, boundary, cluster, and single yellow areas indicate where a plant can be placed (active areas). Circles represent plant positions with their radius. (e), (f): lots and buildings are represented as polygons: $P_L$,$P^{1}_{b}$,$P^{2}_{b}$. To identify the area within a lot that can be used to place vegetation we subtract building polygons from the lot polygon, resulting in the final polygon $P$ (g).}
    \label{fig:placement_strategies}
     \vspace{-4mm}
\end{figure}

As shown in Fig.~\ref{fig:zones} we further assume that a city layout is organized as individual lots, where each lot represents a property that may be occupied by a building. Given a lot and its zone type we then define a PPM that places vegetation individually into each lot. 

\section{Procedural Urban Vegetation} \label{sec:pum}
Vegetation for an urban landscape is generated in two steps: first we apply a PPM to seed plants individually for each zone according to their functional types. After the plants have been planted, we use a developmental model that dynamically grows the plants in their locations while interacting with the surrounding environment. This allows plant adaptation to its environment such as bending and shedding of branches due to the competition for resources, resulting in vegetation with high visual fidelity.

\subsection{Procedural Placement Models - PPMs}\label{sec:ppm}
Our goal is to model the variance of plant morphology and plant placement across municipality zones to realistically distribute vegetation in an urban landscape. Defining and parameterizing rules for obtaining plausible plant positions, while at the same time adhering to urban features such as buildings and streets, is intractable. Therefore, we factorize the problem 
into specifying placement models for the different zones from Sect.~\ref{sec:zones} (industrial, commercial, residential, street, and other) and for each individual lot (Fig.~\ref{fig:placement_strategies}). 

This factorization allows us to define a more manageable parameterization along with placement strategies for the different zones. 
Each placement model defines a concise strategy to place vegetation into a single lot. For example, we have models to place vegetation randomly, along the edges of a lot, equidistantly, etc. Moreover, we define the PPMs in a context-sensitive way. This means to maintain a global appearance, a PPM can query adjacent lots to adjust its parameters (\eg the distance between trees alongside a road in one lot should be the same in the neighboring lot). 
A PPM is defined as the tuple 
\begin{equation}
\mathcal{M} = \left<\mathcal{S}_g, \mathcal{P}_p, \mathcal{P}_s\right>, \label{eqn:ppm}
\end{equation}
where $\mathcal{S}_g$ is a function implementing a \textit{placement strategy} (rules) with $g\in\{R, B, C, E, S, I\}$ (see Sect.~\ref{sec:g} and Tab.~\ref{tab:strategie_vs_param}), $\mathcal{P}_p$ is a set of \textit{positional} parameters to define the placement of plants, and $\mathcal{P}_s$ is a set of \textit{structural} parameters for the morphological appearance of vegetation within the lot. 

\begin{figure}[t]
    \centering
    \includegraphics[width=\linewidth]{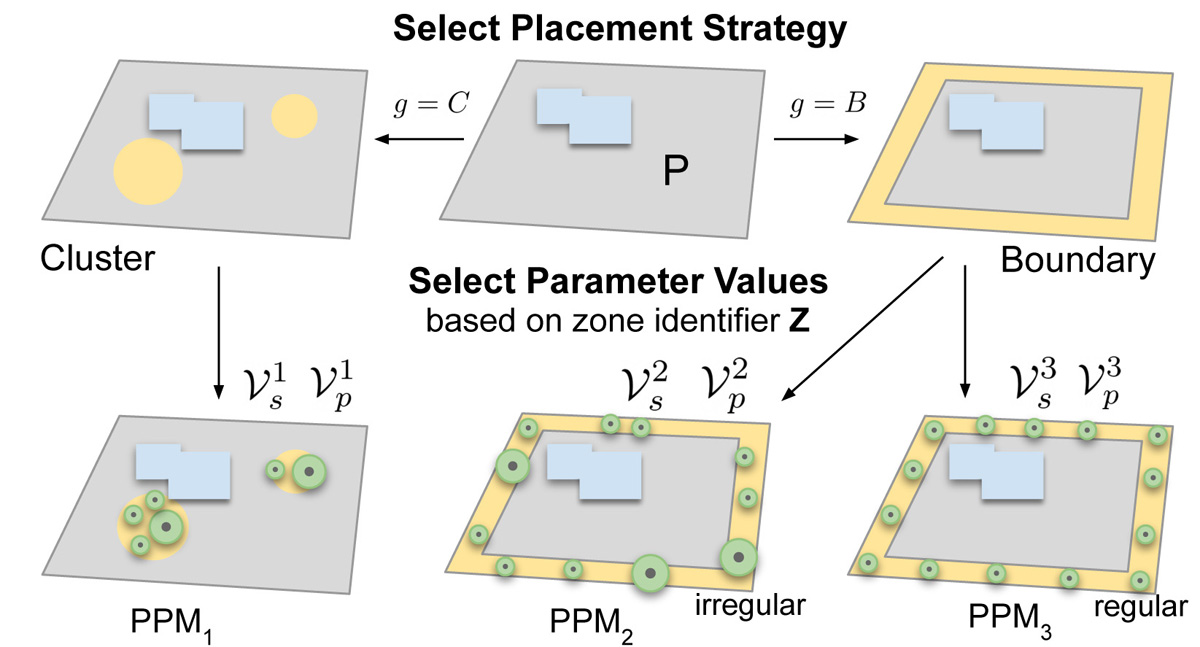}\\
    \vspace{-4mm}
    \caption{Given a lot, we use a placement strategy to define the placement of vegetation. The zone identifier $Z$ is used to select parameter values for structural $\mathcal{V}_s$ and positional parameters $\mathcal{V}_p$. Together, strategies and parameters allow us to generate vegetation with globally similar appearance depending on the municipality zones within a city.}
    \vspace{-4mm}
    \label{fig:strategy_params}
\end{figure}

Lots and buildings are defined as 2D polygons (possibly concave and with holes): $P_{L} = \{V_{L}, E_{L}\}$, $P_{H} = \{V_{H}, E_{H}\}$, where $V_L$ and $V_H$ denote the vertices of a lot ($L$) and buildings~($H$) and $E_L$ and $E_H$ the edges of the polygon for lot and building, respectively. A~lot can include multiple buildings (or other structures): $U=\{P^{i}_{H}\}$. The polygon $P=P_{L} - P^{i}_{b}, \forall P^{i}_{b} \in U$, defines the area of a lot that can be covered by vegetation (Fig.~\ref{fig:placement_strategies}, e-g); the PPM only places vegetation within the geometric shape of the polygon $P$. A~set of plant positions for a single lot is then generated as 
\begin{equation}
 \mathcal{X} = \mathcal{S}_g(\mathcal{V}_p, \mathcal{V}_s, P, Z, \mathcal{K}\label{eqn:seeds}),
\end{equation}
where $\mathcal{V}_p$ and $\mathcal{V}_s$ denote the parameter values for positional~$\mathcal{P}_p$ and structural $\mathcal{P}_s$ parameters, $P$ is the polygon of a single lot, $Z$ is a zone identifier, and $\mathcal{K}$ is the context of a lot. We use $Z$ to select parameter values for a lot. For example, a residential and a commercial lot may use the same strategy (\eg boundary), but differ in their parameter values (\eg different species are used).This is illustrated in Fig.~\ref{fig:strategy_params}. Generating vegetation with the same value for $Z$ produces a uniform appearance  (the same settings are used for every lot), while varying $Z$ with the functional zones generates a diverse and yet coherent appearance. Put differently, $Z$ allows us to control the placement of vegetation on a global scale. Finally, we use~$\mathcal{K}$ to modify the input parameters according to the neighbors of a lot to allow for consistent global appearance as detailed in Sect.~\ref{sec:env}. 

To summarize: a PPM defines a placement strategy along with structural and positional parameters for populating single lots. Varying the values of these parameters generates different plant positions within the constraints of the strategy at a local scale, while changing the parameters jointly -- \eg based on zoning types -- allows us to vary vegetation at a more global scale. 

\begin{figure}[t]
    \centering
    \includegraphics[width=\linewidth]{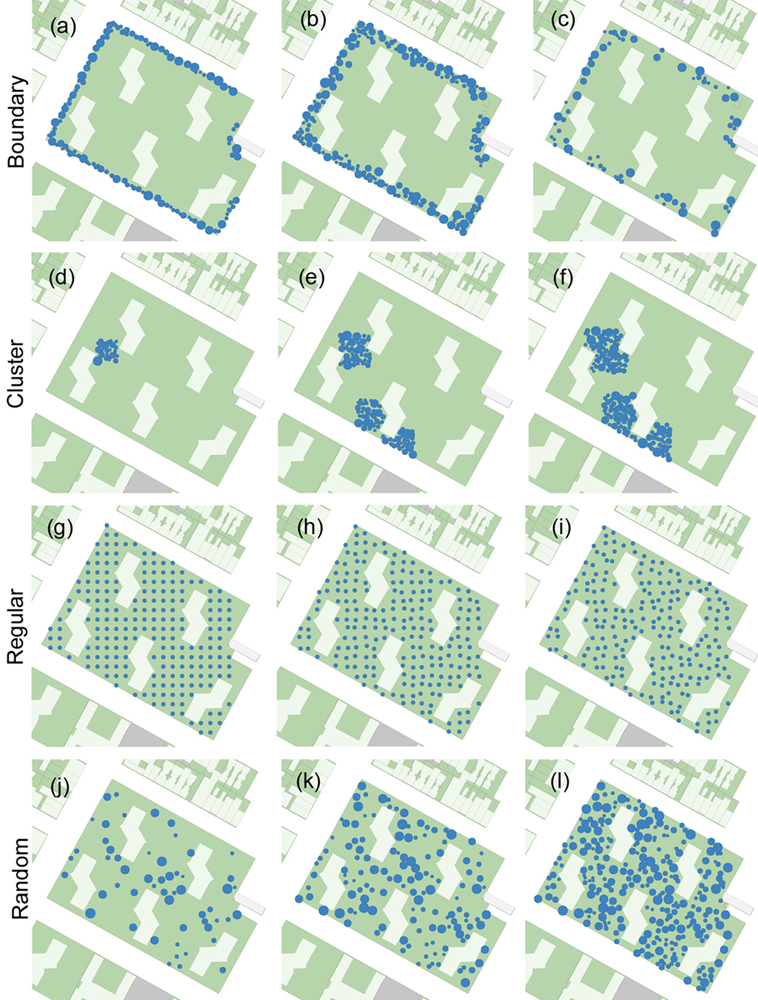}\\
    \vspace{-2mm}
    \caption{Variations of positional parameters on a single lot with different placement strategies. (a)-(c): strategy \textit{boundary} with narrow (a) and wide (b) boundary size, and less density (c). (d)-(f): strategy \textit{cluster} with a single cluster (d) and multiple clusters (e) of different sizes (f). (g)-(i): strategy \textit{regular} with no (g), medium (h), and high (i) jitter. (j)-(l): strategy \textit{random} with low (j), medium (k), and high (l) density. 
    }
    \vspace{-5mm}
    \label{fig:positional_params}
\end{figure}

\subsection{Placement Strategies}\label{sec:g}
A placement strategy $g\in\{R, B, C, E, S, I\}$ defines rules for placing the plants and how the parameters are used. Specifically, we define the following strategies: (1)~\textit{random} placement within an entire lot, (2)~along lot \textit{boundaries}, (3)~  \textit{clustered} distribution, (4)~\textit{equidistant} along the medial axis of a lot, (5)~placement of a \textit{single} plant, and (6)~ \textit{regular} placement within an entire lot.

To implement the different strategies, we compute active areas within each lot that define where the vegetation can be placed. For the strategies \textit{random} and \textit{single} the entire lot polygon is used, while for the strategies \textit{boundary} and \textit{cluster} we define active areas within the polygon; \ie we define a boundary along the edge of the polygon towards its center for \textit{boundary} and a circular area around a randomly selected point within the polygon for \textit{cluster}. For \textit{equidistant}, we compute the medial axis of the polygon and then generate equidistant plant positions along the axis. The strategy \textit{single} defines the random placement of a single plant within the entire lot. Finally, for \textit{regular} we compute a lot-aligned lattice and place plants at the center of each cell. Fig. \ref{fig:positional_params} shows four of our six placement strategies and their parameter  variations.

\begin{table}[t]
\footnotesize
\centering
\caption{Positional and Structural Parameters for PPMs}
\vspace{-2mm}
\label{tab:param_table}
\scalebox{1.0}{
\begin{tabular}{c|clc}
\multicolumn{2}{c}{\textbf{Parameters}} & \textbf{Meaning} & \textbf{Range/Dimensions}\\ \hline
\multirow{11}{*}{\rotatebox[origin=c]{90}{\textbf{Positional}}}
& $\mu$    & Tree envelope mean      & [1m - 10m] \\
&$\sigma$  & Tree envelope variance   & [0.1 - 2] \\
&$\tau$    & Vegetation density       & [0-1] \\    
&$\beta$   & Boundary size            & [0m - 5m] \\    
&$\kappa$  & Cluster radius       & [1m - 20m] \\  
&$\pi$     & Max number clusters      & [0 - 5] \\ 
&$\omega$  & Regularity grid size     & [5m - 50m] \\  
&$\psi$    & Regularity jitter        & [0 - 1] \\
&$\eta$    & Regularity orientation        & [0 - 180\textdegree] \\
&$\delta$  & Equidistant spacing     & [0m - 10m] \\
&$\xi$  & Radius of context     & [0m - 300m] \\
\hline 
\multirow{5}{*}{\rotatebox[origin=c]{90}{\textbf{Structural}}}&$\alpha$  & Max plant age              & [0 - 100 years] \\ 
&$\rho$          & Tree vs shrub ratio      & [0 - 1] \\    
&$\theta$        & Species diversity        & [0 - 1] \\    
&$\gamma$        & Pruning factor           & [0 - 1] \\    
&$\lambda$       & Num. species              & [1 - 10] \\    
\hline
\end{tabular}
}
\vspace{3mm}
\footnotesize
\centering
\caption{Placement Strategies and used Positional Parameters.}
\vspace{-2mm}
\label{tab:strategie_vs_param}
\scalebox{0.95}{
\begin{tabular}{l|c|ccccccccccc}
\textbf{Strategy}     &  \textbf{Symbol} & \textbf{$\mu$ }     & \textbf{$\sigma$}    & \textbf{$\tau$} & \textbf{$\beta$} & \textbf{$\kappa$} & \textbf{$\pi$} & \textbf{$\omega$} & \textbf{$\psi$} & \textbf{$\eta$} & \textbf{$\delta$} & \textbf{$\xi$} \\ 
\hline
Random       &  R      & \checkmark & \checkmark  & \checkmark & & & & & & & & \checkmark\\    
Boundary     &  B      & \checkmark & \checkmark  & \checkmark & \checkmark & & & & & & & \checkmark\\    
Cluster      &  C      & \checkmark & \checkmark  & \checkmark & & \checkmark & \checkmark & & & & & \checkmark \\    
Equidistant  &  E      & \checkmark & \checkmark  & \checkmark & & & & & & & \checkmark & \checkmark\\    
Single       &  S      & \checkmark & \checkmark  & \checkmark & & & & & & & & \checkmark\\
Regular      &  I      & \checkmark & \checkmark  & \checkmark & & & & \checkmark & \checkmark & \checkmark & & \checkmark\\ 
\hline    
\end{tabular}
}
\end{table} 

\subsection{Positional Parameters} 
The placement strategies are parameterized by the positional parameters shown in Tab.~\ref{tab:param_table}. For the placement strategies \textit{random}, \textit{boundary}, and \textit{cluster} we use a method called Variable Radii Poisson-Disk Sampling~\cite{mitchel2012} to generate plant positions within active areas of a lot. More specifically, we are interested in generating a set of points $\mathcal{X}$ with spatially varying point density. A new position sample $y$ is assigned a radius $r(y) : \Omega \rightarrow \mathcal{N}(\mu, \sigma)$, where $\mathcal{N}$ denotes a normal distribution with  mean $\mu$ and variance~$\sigma$. The new position sample $y$ is accepted and added to the set if $|y-x| \geq r(x)+r(y) \forall x \in \mathcal{X}$. 

For the \textit{boundary} placement strategy we further define the boundary size as parameter $\beta$. We use $\beta$ to define an area along the normal of the edge of a polygon towards its center. To implement the \textit{cluster} strategy, we randomly sample a number of points in a lot and define the cluster area as a circle with a radius $\kappa$. A lot can have a variable number of clusters with the maximum number defined by $\pi$. For both strategies, \textit{boundary} and \textit{cluster}, we first compute the active regions (boundary, cluster circles) before generating sample positions. For the strategy \textit{single} we sample one position somewhere in the lot. 

To allow for more regular vegetation placement we define the strategies \textit{regular} and \textit{equidistant}. For strategy \textit{regular}, we compute a regular lattice based on the bounding box of a lot and define the size of cells with $\omega$ and their orientation with $\eta$. We optionally jitter the positions using~$\psi$ within each cell. To implement strategy \textit{equidistant} we first compute the medial axis of the lot polygon $P_{L}$~\cite{Choi1997NewAF} and then use it to equidistantly place plant positions along the axis based on the distance parameter $\delta$. We model the \textit{density of vegetation} for all placement strategies by defining the parameter $\tau$, which deactivates position samples in~$\mathcal{X}$. A value of $\tau=1$ activates all samples, while a value of $\tau \leq 1$ randomly deactivates them until all samples are deactivated ($\tau = 0$). Finally, we define the radius $\xi$ for the context $\mathcal{K}$ of a lot. The context is defined as the adjacent lots and we use it to model context-sensitivity (see Sec.~\ref{sec:env}). 

\begin{figure}[t]
    \centering
    \includegraphics[width=0.49\linewidth]{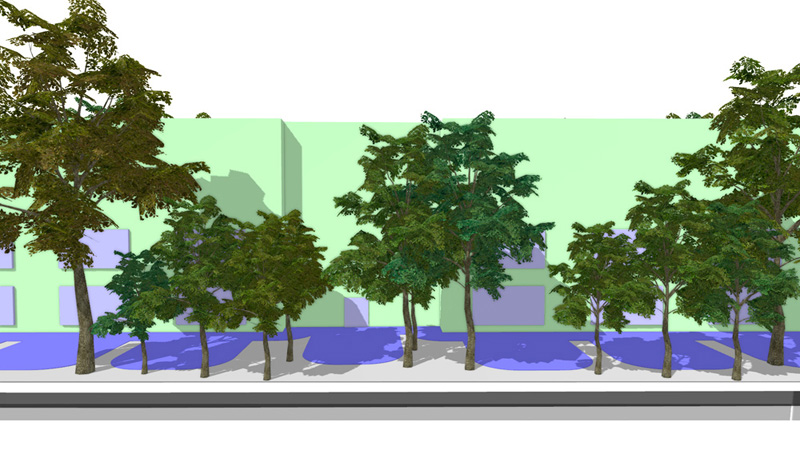}
    \includegraphics[width=0.49\linewidth]{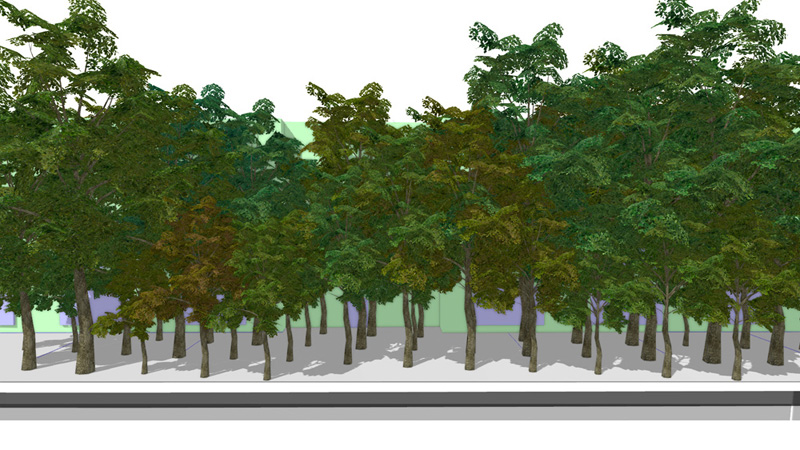}
    \includegraphics[width=0.49\linewidth]{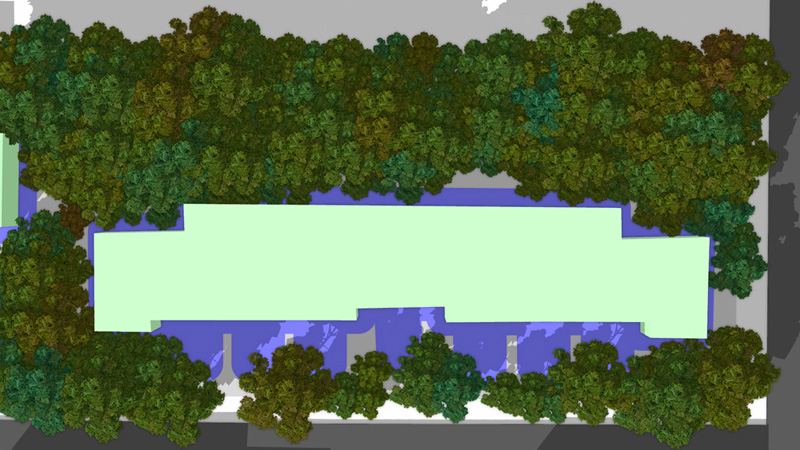}
    \includegraphics[width=0.49\linewidth]{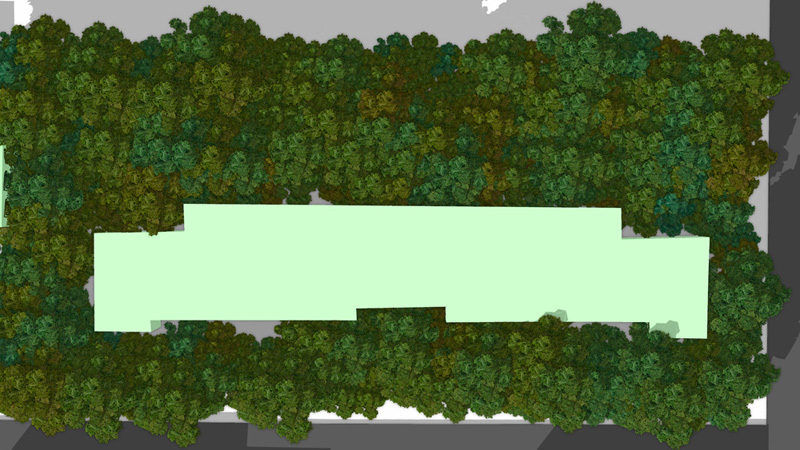}
    \caption{Left: the building envelope (blue) defines a zone where plants cannot be planted to avoid proximity to walls and blockage of door and windows. Right: plant placement without considering the building envelope.}
    \label{fig:envelope}
\end{figure}

The positional placement of plants should also account for the planting rules discussed in Sect.~\ref{sec:zones} \ie trees should not be too close to buildings and should not obstruct doors and windows. We adopted the concept of building envelopes~\cite{miller2015urban} that defines the distances from the buildings. Moreover, we extend the envelope in front of doors and windows to avoid their blockage. An example in Fig.~\ref{fig:envelope} shows the effect of using the building envelope.

Tab.~\ref{tab:param_table} summarizes all positional parameters along with their ranges, while Tab.~\ref{tab:strategie_vs_param} shows our placement strategies along with the positional parameters that are used for each of them. Examples of changing the values of positional parameters are shown in Fig.~\ref{fig:positional_params}.

\subsection{Structural Parameters}

\begin{figure*}[t]
    \centering
    \includegraphics[width=\linewidth]{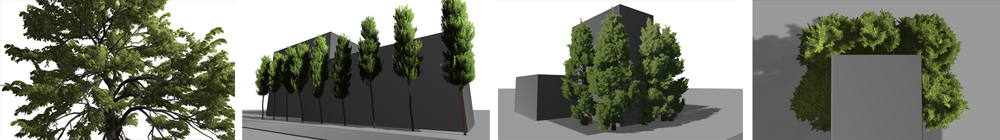}\\
    \caption{Closeup renderings of detailed plant models. As our method relies on a environmentally-sensitive developmental model, we can produce detailed branch geometry that adapts to buildings and other plants.}
    \label{fig:closeup}
\end{figure*}
\begin{figure}[t]
    \centering
    \includegraphics[width=\linewidth]{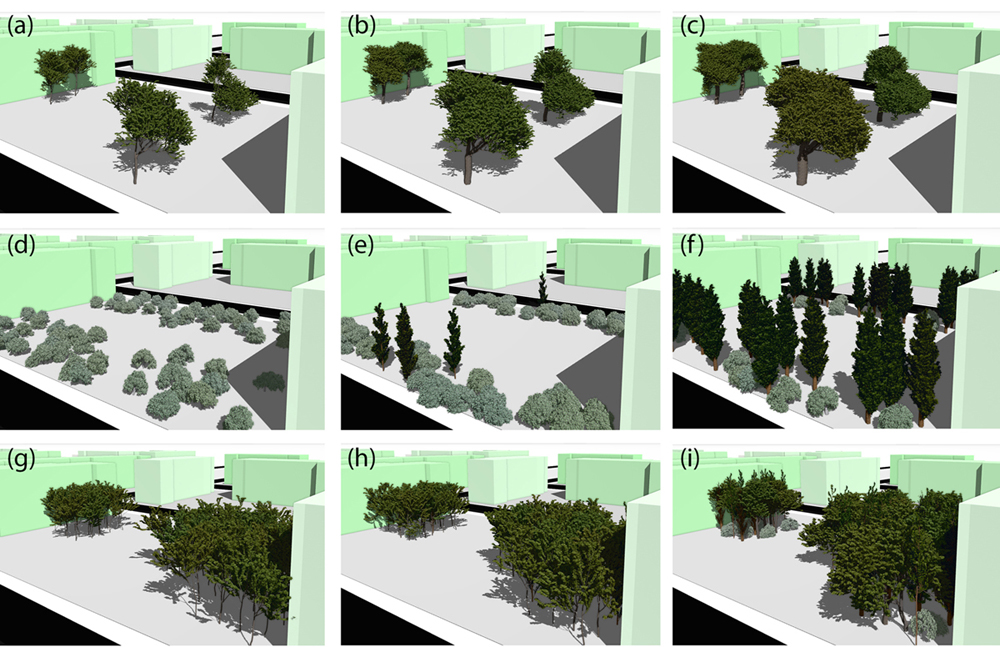}\\
    \vspace{-3mm}
    \caption{Variations of structural parameters. Top row: variations of age parameter from young (left) to old (right). Middle row: changes of tree to shrub ratio from only shrubs, to mostly trees. Bottom row: variations of species diversity from a single species (left) to multiple species (right).}
    
    \label{fig:structural}
\vspace{1mm}
    \centering
    \includegraphics[width=1.0\linewidth]{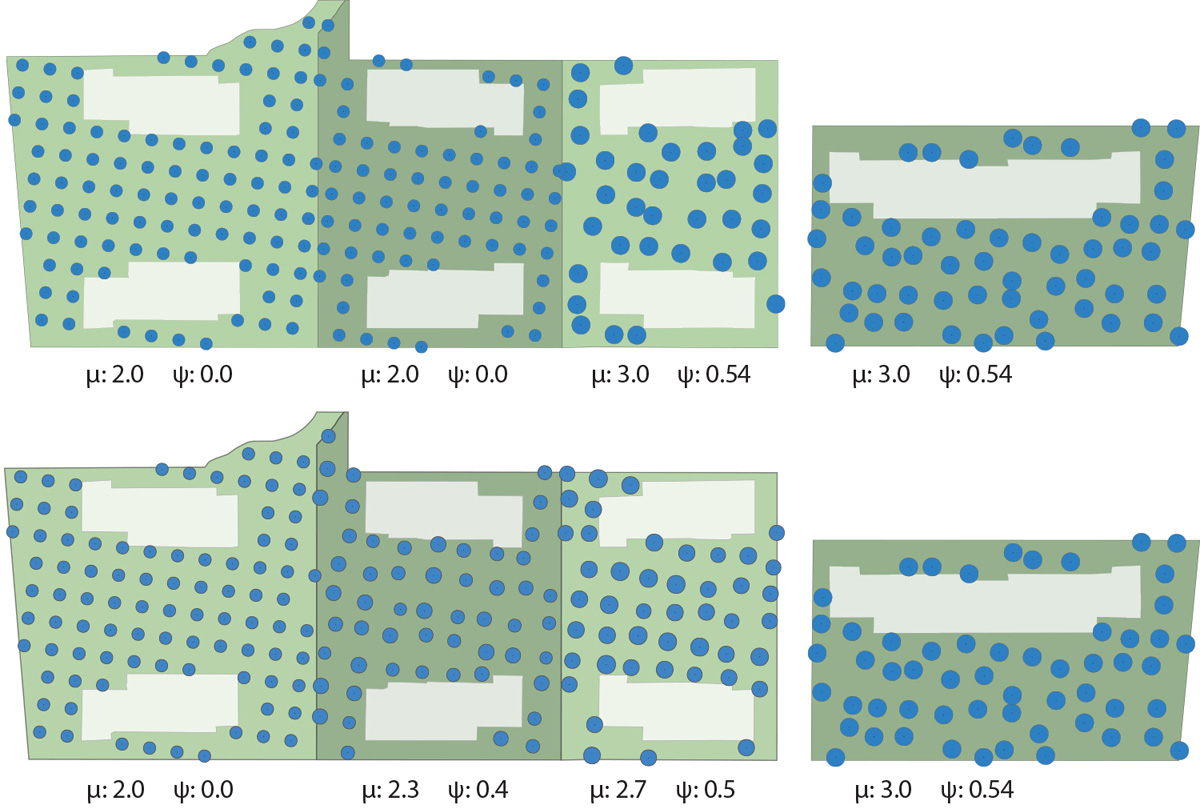} \\
   \vspace{-3mm}
    \caption{Context-sensitivity: we calibrate the parameter values of a lot with those of adjacent lots (context). Here we show two lot configurations with regular placement strategy and variations over the parameters $\mu$ and~$\sigma$. For the lots shown in the top row context-sensitivity is turned off and plant placement changes abruptly from one lot to another, while for the bottom row we show context-sensitivity across lots and the resulting calibration of parameters (context radius: \textbf{$\xi=180m$}). }
\label{fig:context}
\vspace{-2mm}
\end{figure}

We define \textit{structural parameters} to model the morphology of individual trees as well as the plant population within a lot. Based on the computed plant positions we define a plant seed as the tuple  
\begin{equation}
\mathcal{T} = \left<\ p, \alpha, \phi, \gamma\right>, \label{eqn:morph}
\end{equation}
where $p \in \mathcal{X}$ is the plant position, $\alpha$ its maximum age, $\phi$ denotes a species identifier, and $\gamma$ is a pruning factor. To generate branching structures we grow a plant with a developmental model (see Sec.~\ref{sec:dev_model}) and jointly simulate its growth with all other plants in a lot.

We define a number of species ($n=10$) for the whole urban landscape by selecting parameter values for our developmental model~\cite{Palubicki:2009:STM}. We then use the species identifier $\phi$ to associate one of the species to a seed. We further control this selection by using the parameter~$\rho$, which defines the tree vs. shrub ratio in a lot. A value of $\rho=1$ assigns all seeds tall-growing species, while a value of $\rho=0$ only associates short growing ones. 

To vary the number of used species in a lot we use the parameter~$\theta$. We randomly select one of the species as the dominant species in a lot and use $\theta$ as a ratio to control the number of seeds associated with the dominant species and all other available species. A value of $\theta=0.5$ sets half of the available seeds to the dominant species and the other half with randomly selected ones. 

Finally, we may prune a plant by a bounding volume for the tree crown of a fully developed model. This allows us to generate a more organized appearance of vegetation, \eg along avenues or highways. Branches that reach out of the volume are cut off. We scale this volume by $\gamma$; a value of $\gamma=1$ will leave a plant unpruned, while a  value of $\gamma\leq 1$ scales the bounding volume and therefore results in a pruned plant. After pruning, we again simulate the plant growth to develop smaller branches and leaves. Fig.~\ref{fig:pruned_trees} shows an example of the pruning of trees, other variations of structural parameters are shown in Fig.~\ref{fig:structural}. 

\subsection{Context-Sensitive Rules}\label{sec:env}
So far, lots have been treated as individual units without any mutual relationship. However, each lot has its context that are its surrounding roads and neighboring lots. The neighbors often share similar planting rules that are provided by the applying municipal tree ordinances~\cite{grey1995urban,miller2015urban}. We want to define planting rules in a way that would consider the context.

While context-sensitive plant seeding has not been addressed before, there is a body of related work on the environmental sensitivity of individual plants that is closely related to a plant's ability to adapt to varying conditions \eg it may bend its branches against gravity (gravitropism) or grow towards the brightest spot (phototropism). A plant optimizes different functions by using this plasticity. Context-sensitivity can be proceduralized as context-dependency, for example by using environmental query modules in Open L-systems~\cite{Mech:96:SIGG}. Fetching the context values is a two-pass method: first, the context is queried, then the values are interpreted by the procedural system.

Inspired by this previous work, we generalize the context dependency to PPM. Let us recall that each PPM from Eqn.~(\ref{eqn:ppm}) has associated a placement strategy~$\mathcal{S}_g$ and two sets of parameters~$\mathcal{P}_p$ and~$\mathcal{P}_s$. Each lot has a set of parameter values from Eqn.~(\ref{eqn:seeds}) $\mathcal{V}_p$ and~$\mathcal{V}_s$. Moreover, it considers the context (\ie the neighborhood)~$\mathcal{K}$ of the lot that is being populated with plant positions Eqn.~(\ref{eqn:seeds}). 

Let us denote a particular lot of interest~$L$ and its parameter values as~$\mathcal{V}^L$. In the following text, we will omit the lower index $s$ and $p$, because the parameters are calculated in the same way. The context is the set of lots within radius~$\xi$ centered on the lot~$L$ and weighted by a 2D Gaussian. The values of the corresponding parameters (see Tab.~\ref{tab:param_table}) of the neighbors and the lot~$L$ are weighted according to the distance resulting in a context-updated parameter set $\tilde{\mathcal{V}^L}$ as:
\begin{equation}
\tilde{\mathcal{V}^L}=\sum_{\forall\mathcal{V^K}\in\mathcal{K}} w\left(d(L, L^{K}) \right) \mathcal{V}^K,\label{eqn:vtilde}
\end{equation}
where $w\left(d(L, L^K)\right)$ is the Gaussian-weighted distance of the center of the lot~$L$ from the center of the lot~$L^K$ within the investigated context and $\mathcal{V}^K$ are the values of the parameters of the lot~$L^K$. The updated parameter values~$\tilde{\mathcal{V}^L}$ are then used for the PPM.

Note that this process can be considered as a diffusion of the parameters within radius~$\xi$. Also, to avoid a race condition when one lot serves as a context of another one and vice versa, we calculate the context-updated parameters~$\tilde{\mathcal{V}^L}$ into a different map. In this way, the calculation does not depend on the order of the lot selection and can be also done in parallel. Note that if we would apply Eqn.~(\ref{eqn:vtilde}) multiple times, the values of the parameters would be smoothed out into an average over the entire layout.

An example in Fig.~\ref{fig:context} shows the effect of using context sensitivity on a regular placement of trees. The first row shows two lots with regular tree placement with an abrupt change to a random placement in neighboring lots that is smoothed out into a semi-random transition when the context is used (bottom row).

\subsection{Developmental Plant Model} \label{sec:dev_model}
After generating plant positions, we jointly grow the plants in the computed locations of a single lot. Our developmental model is based on the work of Palubicki et al.~\shortcite{Palubicki:2009:STM}; a tree is a modular system (leaves, buds, stems, and internodes). An internode is a plant stem between two or more leaves and a tree is composed of a succession of internodes. 

The primary plant development is controlled by the expansion of buds that are either apical (terminal) or lateral (axial). Branches expand at their tips by expanding their apical buds or on sides by growing lateral buds. Buds use signaling by the growth hormone Auxin to prevent overgrowth and to control apical dominance~\cite{Kebrom17Buds}. Secondary plant development (cambial growth) is the thickening of a tree trunk and branches~\cite{Kratt15} simulated by expanding their radii using da Vinci's rule (see~\cite{minamino2014tree} for a discussion). 

Trees compete for space by seeking light (phototropism) and avoiding collisions and overcrowding. Many different algorithms have been implemented to capture plant competition for resources (see~\cite{Runions07,Mech:96:SIGG} and~\cite{Pirk16STAR} for an overview). We use the approach of~\cite{Runions07} later extended by~\cite{Palubicki:2009:STM} that uses space occupation by randomly scattered particles that attract growing branches. We also simulate phototropism by computing the illumination of buds and bending the growth direction towards the brightest spot visible from a bud. Apical control and branching parameters are simulated by using the growth model from~\cite{Stava2010} with the set of parameters.

\begin{figure}[t]
    \centering
    \includegraphics[width=\linewidth]{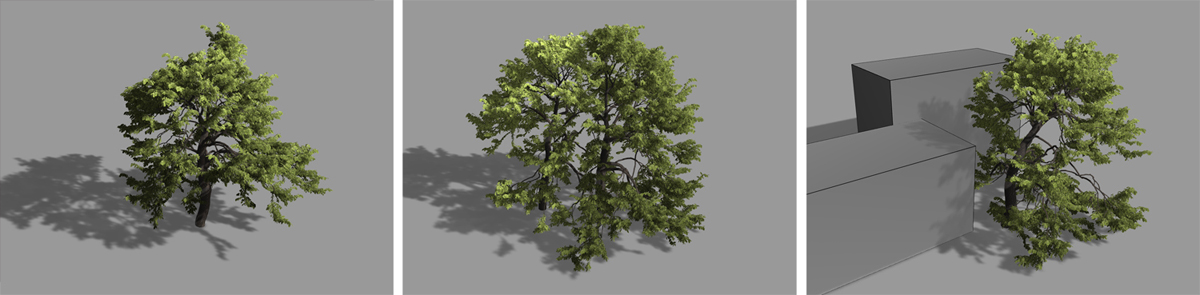}\\
    \includegraphics[width=\linewidth]{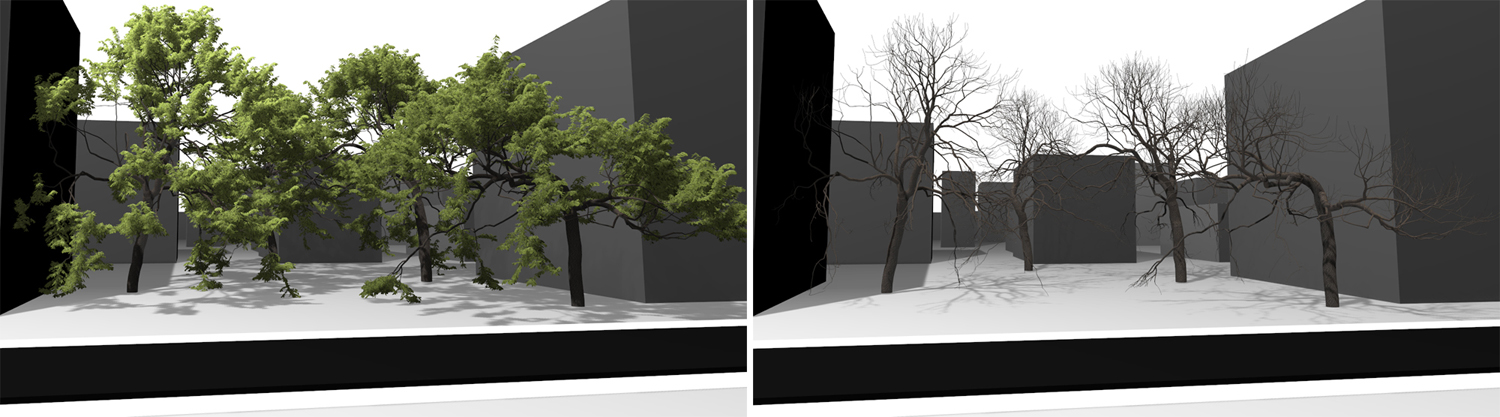}\\
    \vspace{-3mm}
    \caption{Top row: a tree grown in different environmental conditions. From left to right: alone, together with another tree, and close to a set of buildings. Bottom row: the growth response of a group of trees in an urban environment generates complex and unique branching structures.}
    \label{fig:individual_growth}
\vspace{1mm}
    \centering
    \includegraphics[width=\linewidth]{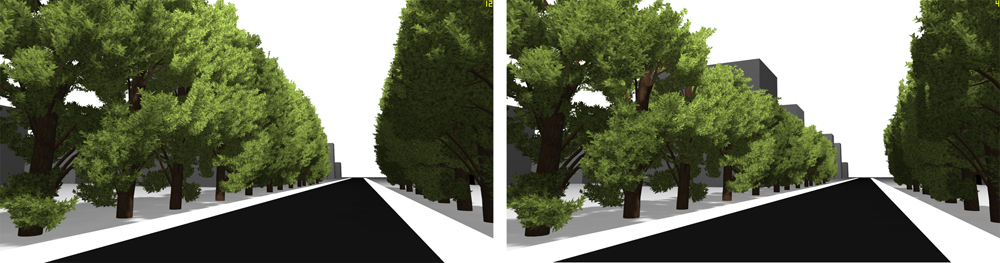}\\
    \vspace{-3mm}
    \caption{Pruning of branches allows for the adjustment and organization of tree form. Here trees along a street are severely pruned to form a hedge ($\gamma=0.7$).}
    \label{fig:pruned_trees}
\end{figure}
\section{Learning Vegetation Placement} \label{sec:learning}
Learning plant positions directly from image data is a challenging problem that cannot be easily addressed by existing neural network architectures or other methods. To obtain plant positions in an  end-to-end manner, a network would have to either output a variable number of plant positions, or operate on a fixed size domain, such as an image. The latter requires to obtain plant positions as a post-processing step, which is error-prone. Furthermore, generating ground truth data pairs of satellite images and plant positions (\eg GPS coordinates) for training a neural network is challenging (see Sect.\ref{sect:discussion} for a discussion). Moreover, an end-to-end deep learning-based system would sacrifice the in-depth understanding of the underlying mechanisms and would not allow for low-level control that is needed in interactive editing. 

Therefore, to recover the placement and appearance of real urban landscapes we aim to learn the distribution of plants in the parameter space of our positional parameters. This has the advantage that our above-defined PPMs act as a prior, which helps to regularize the training of our network and in turn to generate plausible plant positions. Furthermore, learning the parameters of a procedural model maps images to a set of comprehensible and intuitive parameters, which provides an efficient way to further edit plant placements.

\subsection{Learning Plant Placements}
\begin{figure}[t]
    \centering
    \includegraphics[width=\linewidth]{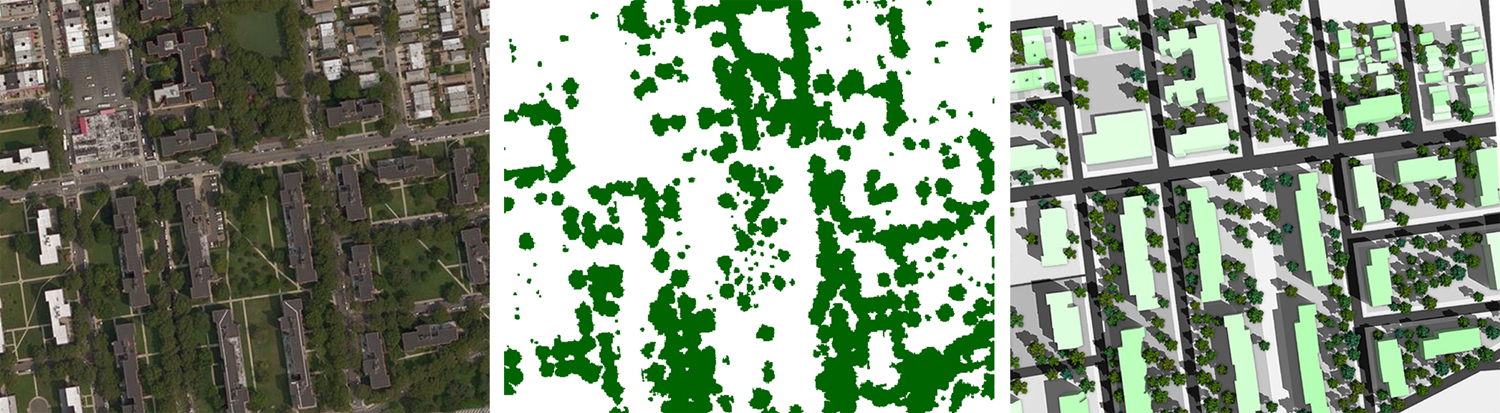}\\
    \caption{Vegetation placement based on real data: we use vegetation cover-age maps (middle) to identify active regions for individual lots and populate them with our PPMs. This allows us to generate plant distributions~(right) similar to what can be observed in satellite images (left). 
    }
    \vspace{5mm}
    \label{fig:veg_placement_coverage}

    \centering
    \includegraphics[width=\linewidth]{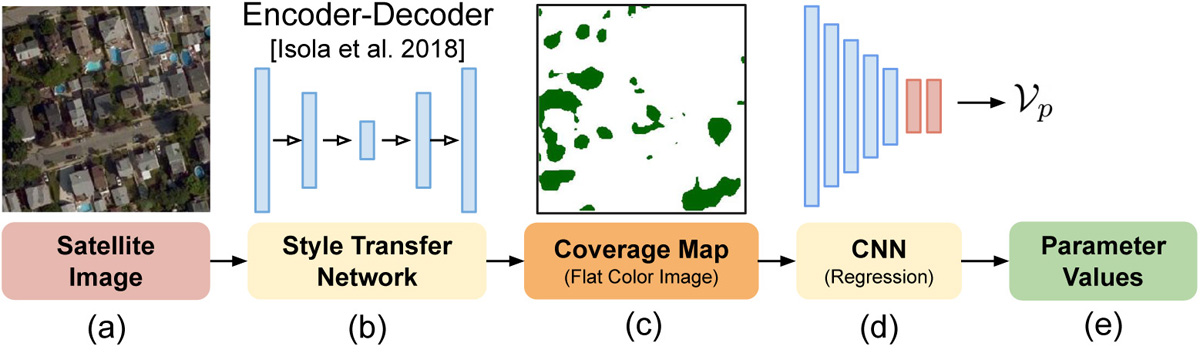}\\
    \caption{Neural network pipeline: we use a style-transfer network (b) trained on data pairs from NYCOpenData~\shortcite{NYCData2019} to convert satellite images (a) to coverage maps (c). To learn parameter values for our PPMs (for which no ground truth data for satellite images exist) we generate pairs of coverage maps and parameter values with the PPMs of our framework. We then train a CNN (d) to obtain parameter values (e) for the estimated coverage maps of the real satellite images.}
    \label{fig:networks}
\end{figure}

We use a two-stage neural network pipeline to learn the parameters of our PPMs: first, we translate satellite images to semantic maps that describe vegetation coverage (Fig.~\ref{fig:networks}, a-c). Second, we learn the positional parameters from coverage maps with a lightweight convolutional neural network  (Fig.~\ref{fig:networks}, d, e). This pipeline has the advantage that we do not need to rely on pairs of satellite images and positional parameters for training, but instead on pairs of coverage maps and positional parameters, which can be generated synthetically with our PPMs. 

To translate satellite images to coverage maps we used a style-transfer deep neural network~\cite{Isola2016ImagetoImageTW}. A coverage map is a flat-colored image where every pixel color is based on whether the corresponding pixel in a satellite image represents vegetation. Coverage maps have less complex visual traits and are similar for real and synthetic data. Therefore, the network is able to learn this transfer. We used pairs of satellite images and coverage maps that are publicly available for some cities~\cite{NYCData2019} to train the style-transfer network to learn coverage maps from satellite images. This allows us to obtain coverage maps of cities for which coverage data does not exist. 
Fig.~\ref{fig:semantic_maps} (Appx.~\ref{apx:map_images}) shows examples of training data and generated coverage maps.

We then train a neural network to obtain positional parameter values ($\mu, \sigma, \tau, \beta, \kappa, \pi$, see Tab.~\ref{tab:param_table}) from the coverage maps. 
Training is done on synthetically generated pairs of coverage maps and positional parameters obtained from our PPMs. Specifically, we define the generated coverage maps as $q\in\mathcal{Q}$ for which we know the corresponding positional parameters $\mathcal{P}_p \in \mathcal{U}$. The network can thus be defined as 
$$
f(q) : \mathcal{Q} \rightarrow \mathcal{U}.
$$
To summarize: stage one of our pipeline allows us to learn coverage maps from satellite images, which -- in stage two -- allows us to obtain the positional parameters of our PPMs. Together this enables us to generate positions of vegetation for individual lots with similar characteristics as observed in the satellite imagery (\eg plant distance, density, etc.). Once the parameters are generated, we stencil the coverage map with the geometry of each lot and identify areas where we need to place vegetation for a reconstruction. We convert the regions into polygons and then use our PPMs to generate plant positions within the identified areas of a lot (Fig.~\ref{fig:veg_placement_coverage}). As a coverage map defines the areas where vegetation should be placed within a lot, it replaces the placement strategies introduced in Sec.~\ref{sec:g} -- the areas defined by the coverage map are then populated with a random strategy along with the learned positional parameters defined in Tab.~\ref{tab:param_table}.

\subsection{Data and Training}\label{sec:data}
For training the Pix2Pix style-transfer network we rely on the publicly available implementation of the original model implemented in Python. We train the network on 20K pairs of satellite images and coverage maps provided by the NYC Open Data~\shortcite{NYCData2019}. We use the default hyperparameter settings for Pix2Pix~\cite{Isola2016ImagetoImageTW}; the network converged after training for 200 epochs. We then use the network to convert satellite images of urban landscapes to coverage maps. The geometry of single lots is also obtained from the NYC Open Data. Our urban modeling framework operates on longitudinal and latitudinal coordinates, which allows us to register satellite images, lot data, and coverage maps, which in turn enables us to render satellite images and publicly available maps (\eg Open Street Maps) in the same framework. Our regression CNN consists of five convolutional layers (32 units) followed by two dense layers (64 units) with relu activations for all, expect the last layer. We use our PPMs to synthetically generate 21K pairs of (coverage map, positional parameter)-pairs to train the network. To regress the positional parameters we use mean squared error as loss function and are able to achieve 95\% accuracy for predicting the parameters. We use a 80\% -- 20\% split for training and testing data. All results shown in the paper are generated from validation data.

\section{Implementation and Results} \label{sec:results}

Our interactive framework for modeling and rendering urban landscapes was implemented in \texttt{C++} and OpenGL. All results have been generated on an Intel(R) Core i7-7700K, 8x4.2GHz with 32GB RAM, and an NVIDIA Geforce RTX 2080 GPU (12 GB RAM). 

The most demanding online task is the generation of tree geometry. We simplify this by representing trees by their skeletons that are generated on the CPU. We further offload the mesh generation of the branch surfaces into a geometry shader on the GPU. Similarly, leaves are generated as textured quads that are also generated on the fly. Buildings and other structures are rendered as extruded outlines. While we cannot render large plant populations in real-time, our framework allows us to interactively explore placement strategies and parameter settings. To render large scenes (\eg Fig.~\ref{fig:large_scale}) we use a simple level-of-detail scheme that successively replaces tree geometry with billboards and point primitives based on the distance to the camera. Appx.~\ref{appx:params} (Tab.~\ref{tab:param_values_table}) shows parameter values for most figures shown in the paper.

\subsection{Interactive Authoring}
We have shown that PPMs can be used to automatically place vegetation into urban landscapes based on the lot data. The geometry of individual lots can either be obtained from publicly available datasets, or --for synthetically generated urban layouts-- as part of the modeling process. 

\begin{figure}[t]
    \centering
    \includegraphics[width=\linewidth]{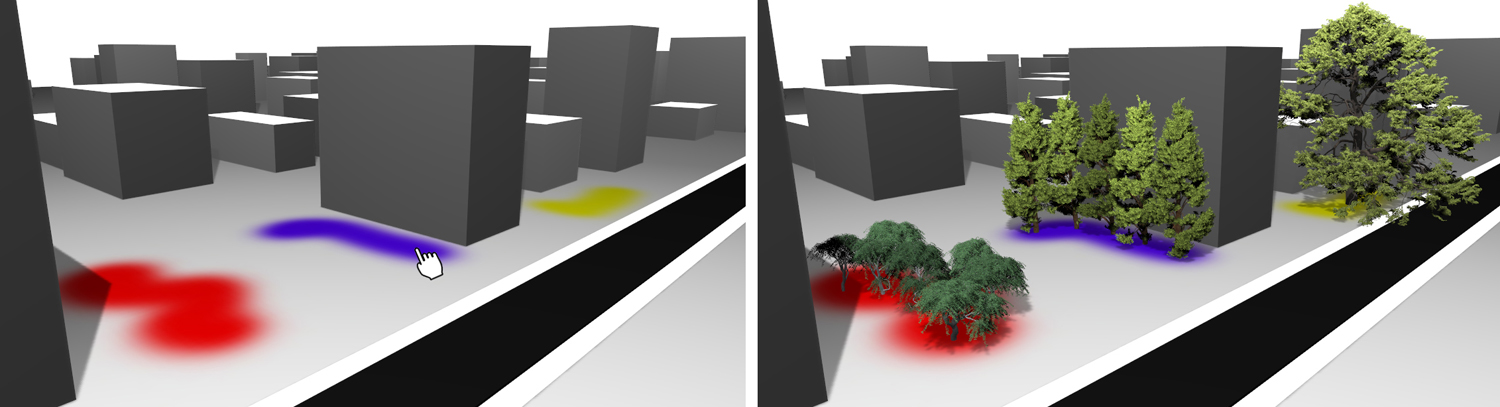}\\
    \vspace{-2mm}
    \caption{A user can interactively sketch placement zones with a brush tool~(left). Each placement zone is converted to a polygon and assigned a placement strategy to grow plants (right). Here we show the strategies medial axis (blue), single (yellow), and random (red).}
    \label{fig:interactive_authoring}
\vspace{3mm}
    \centering
    \includegraphics[width=\linewidth]{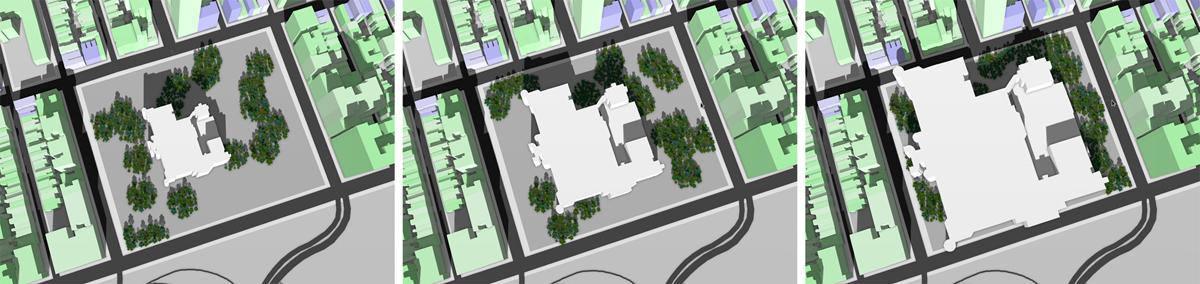}\\
    \vspace{-2mm}
    \caption{The placement of vegetation changes with the size of the active areas within a lot. While the used cluster strategy initially generates plants in the entire lot, transitioning to less available space due to a larger building (white) generates more organized plant positions at the boundary of the lot.}
    \label{fig:growth_of_building}
\end{figure}

However, PPMs operate on a polygon and they were designed with interactive authoring in mind. The user can simply use a brush tool to draw an area on a map. We then convert the sketch to a polygon and assign a PPM. Depending on its placement strategy, the PPM will then generate plant positions according to the geometry of the polygon and its associated placement strategy (Fig.~\ref{fig:interactive_authoring}). Furthermore, a user can directly draw the vegetation coverage for individual lots or polygons. Similar as to learning the coverage maps from satellite images, sketching a coverage map replaces the placement strategy for a lot and the PMM then places plants based on the positional and structural parameters, which provides a convenient way for more nuanced vegetation placement.
 
This process also allows us to generate even more diverse zones if necessary. For example, it is possible to define individual zones for back and front yards, the vegetation along streets, or even parks. A key idea of our approach is to factorize the complexity of defining a complex procedural model into more manageable placement strategies. A PPM only works on a single polygon and generates plant positions for this geometry. This way it is easy to extend our approach by new placement strategies.  

 \begin{figure*}[t]
    \centering
    \includegraphics[width=\linewidth]{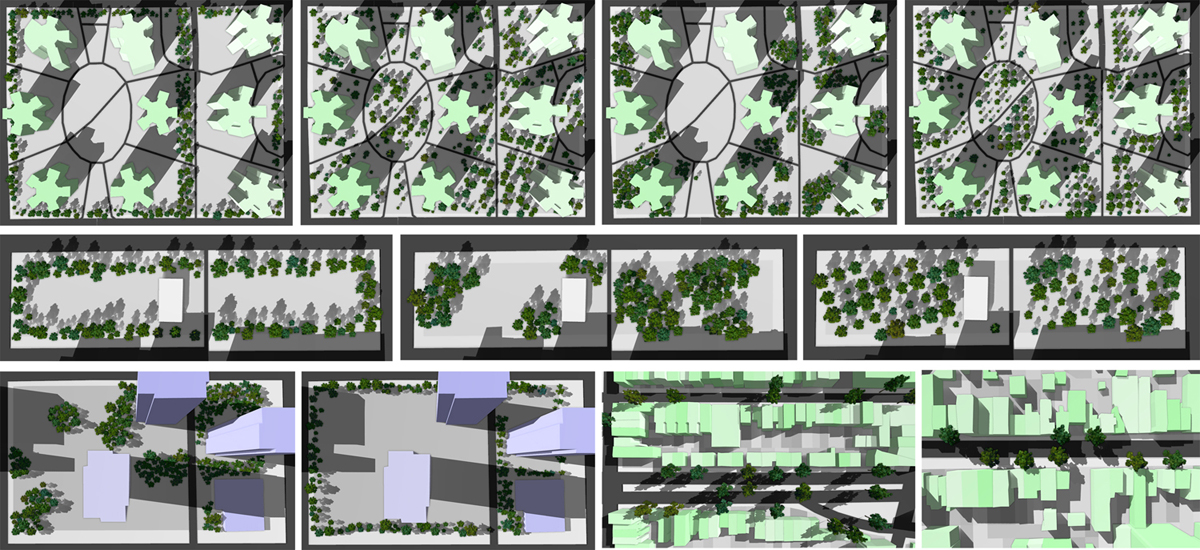}\\
    \caption{Top-down renderings of plant distributions for three municipality zones generated with different placement strategies. Top row: the placement strategies boundary, random, cluster, and regular for a residential lot of buildings. Middle row: the placement strategies boundary, cluster, and regular for the lot of a public park. Bottom row: the placement strategies cluster, boundary for a commercial lot (left) and the placement of trees with medial axis along streets with equidistant spacing set to: $\delta=13m$ (right).}
    \label{fig:vegetation_lots}
 \end{figure*}
 \begin{figure*}[t]
    \centering
    \includegraphics[width=\linewidth]{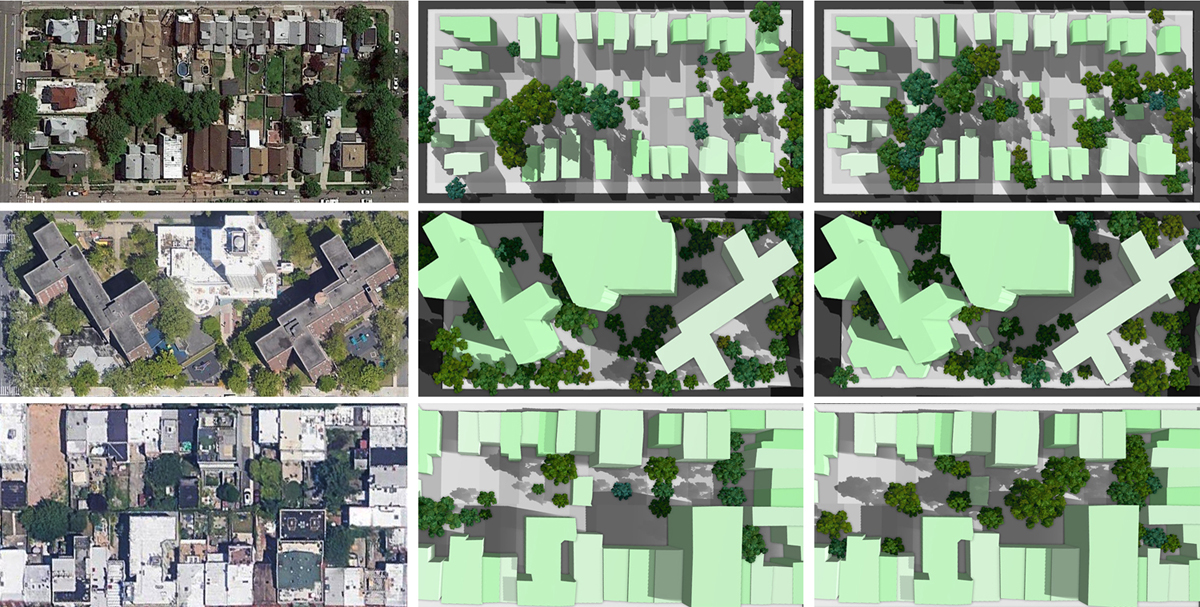}
    \vspace{-4mm}
    \caption{Given a satellite image (left), our method is able to generate similar plant populations (right) as what can be observed in the real scene. To compare the results of our procedural model we manually labeled plants and used their longitude and latitude coordinates to render them at their real positions in our framework (middle). This allows us to evaluate the visual quality of synthetically generated plant positions compared to real plant distributions. }
    \vspace{-2mm}
    \label{fig:reconstruction_manual}
\end{figure*}

\begin{figure*}[t]
    \centering
    \includegraphics[width=\linewidth]{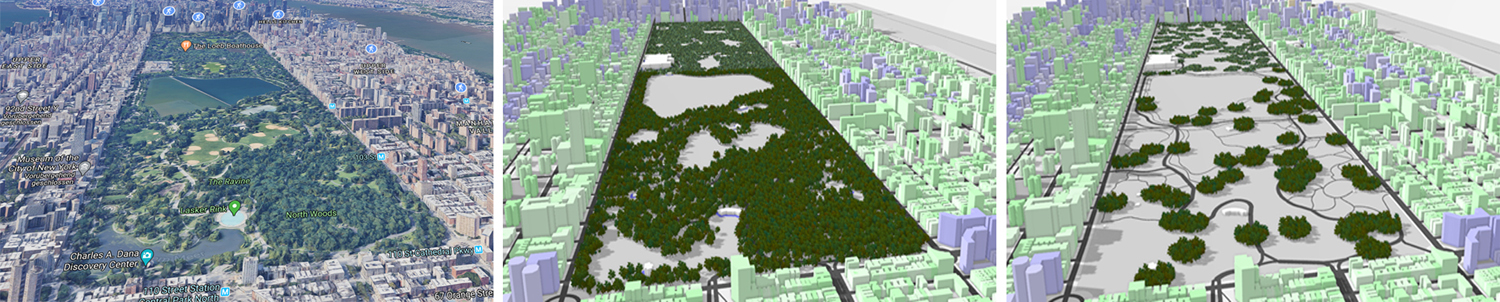}
    \vspace{-7mm}
    \caption{Left: Google maps view of New York (Central Park). Our framework generated two variations of plant placements (middle, right) for an initially empty city model. Middle: 54k plant positions were generated in about 60 seconds. 
    }
    \label{fig:large_scale}
    \vspace{-2mm}
\end{figure*}

\begin{figure*}[t]
    \centering
    \includegraphics[width=\linewidth]{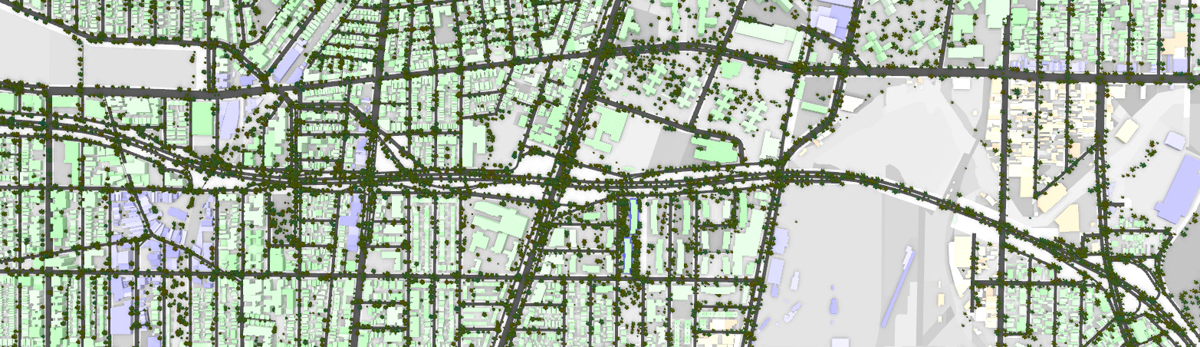}
    \vspace{-7mm}
    \caption{Our framework enables to efficiently place vegetation for large urban areas.}
    \label{fig:large_scale}
    \vspace{-2mm}
\end{figure*}

\subsection{Results}
Figs.~\ref{fig:teaser},~\ref{fig:reconstruction_manual}, and~\ref{fig:large_scale} show perspective and top-down renderings of urban landscapes along with the vegetation generated by our framework. For these results we used coverage maps to reproduce vegetation placement similar to the real scenes. Fig.~\ref{fig:vegetation_lots} shows results where we only used our procedural model, without additional coverage maps. For both cases, the produced plant populations show characteristic visual traits as found in real distributions of vegetation at city-scale. Based on our placement strategies -- in combination with the positional and structural parameters -- we can generate complex patterns of urban vegetation. 

Moreover, we show vegetation placements for the different municipality zones (residential, park, commercial) in Fig.~\ref{fig:vegetation_lots}. Positional parameters allow us to generate planting patterns as commonly found in these areas, while we can also produce structural variations by selecting the number of species, their height, and their age (Fig.~\ref{fig:structural}). Additionally, we can control the pruning of plants to generate more organized plant shapes (Fig.~\ref{fig:pruned_trees}). In an urban setting buildings often shade larger areas. Trees growing in these regions strive to grow out of the shadow toward the light. This interaction of a tree with other trees as well as close-by buildings generates complex and unique branching structures. Figs.~\ref{fig:individual_growth} and~\ref{fig:closeup} show the modeling result of trees grown in varying environmental conditions. 

Finally, Figs.~\ref{fig:interactive_authoring} and~\ref{fig:growth_of_building} show the capabilities of our framework for the interactive authoring of urban landscapes. In Fig.~\ref{fig:interactive_authoring} a user drew regions for vegetation onto the ground of an urban layout; each brush tool was assigned a different placement strategy and set of parameters values. Our method then converted the sketched areas to polygons and applied different PPMs. In Fig.~\ref{fig:growth_of_building} we show how the placement of vegetation changes, when the size of a building on a lot increases. While with a small building there is more space for random plant configurations, the placement transitions to more organized plant positions when the size of the building increases.
\section{Evaluation, Discussion, and Limitations}

To validate point distributions generated with our placement models we ran a user study to evaluate the perceived realism of plant distributions that are generated with our PPMs and real data. Furthermore, we measured the distance of generated and ground truth point sets of plant positions. 

\subsection{Perceptual User Study} \label{lab:user_study}
We generated two sets of images for the user study, one with trees placed by our PPMs and the other based on real data. However, validation against satellite images is difficult, because it is not possible to fully generate the same visual complexity. To avoid this bias, we rendered all images by using our framework (examples are shown in Fig.~\ref{fig:reconstruction_manual}). We identified 28 lots with varying plant placement and produced plant positions by using all placement strategies for these lots and rendered them as top-down images with our framework. We generated the real data by manually identifying plants in satellite images in these lots and marked their positions. We then loaded these positions into our framework and rendered them in the same rendering style for both categories to avoid bias. Furthermore, we chose top-down views for evaluating placement strategies, as this allows us to evaluate the respective distributions of plants.

We then performed a two-alternative force check (2AFC) on the images. We generated pairs of images one being generated from the real data and one from each category of procedural data, leading to the total $28\times{6}=168$ pairs of images. We randomly shuffled their placement (left-right) and their order. The image pairs were shown to 33 users from Mechanical Turk (MT) and we made sure that only MT masters (reliable users) were answering the study. We then asked the users "Which plant distribution looks more realistic (left or right)?" and the user had to choose one image. Each PPM category and real data received multiple rankings from every user. 

Our tests indicate that the PPMs provide results that are perceptually consistent with real-world tree placement. The users selected real as more realistic in 58\% and the procedural placement in 42\%. 

\subsection{Quantitative Evaluation}
To quantitatively evaluate the generated point distributions we measure the Chamfer distance (CD) (e.g. as used in \cite{8099747}) between manually labeled plant positions and the procedurally generated ones. For each point in a point set CD finds the nearest point in the other point set and computes the sum of squared distances. A distance of $0$ would indicate that the two point sets are the same. However, we are not aiming to generate the exact same point set, but one with a similar distribution of positions. We used the manually labeled plant positions of 28 lots (Sect.~\ref{lab:user_study}) and generated random point sets as a baseline. We observe an average distance of 0.16 between manually labeled points (ground truth) and procedurally generated point sets that are supposed to show a similar distribution of positions, and a distance of 0.27 between the ground truth and random point distributions.

While only comparing point sets is not a conclusive metric for evaluating the similarity of procedurally generated and real vegetation, it shows that our model produces plant distributions that are closer to the ground truth positions compared to random positions. 

\subsection{Discussion and Limitations}
\label{sect:discussion}

Our framework allows us to place and simulate vegetation in urban landscapes. To this end, our focus was on generating convincing distributions of plants for synthetic and real city models. Because defining rules for all possible variations of plants in urban landscapes is intractable, we factorized the problem of placing plants into a number of placement strategies. Each strategy provides a concise set of rules along with parameters to describe the positional and structural properties of vegetation within individual lots. Together, placement strategies and parameters allow us to generate realistic distributions of plants within the functional zones of an urban layout. In addition, we use a state-of-the-art developmental model for plants to simulate their environmental response. 

We generate distributions of vegetation that resemble what can be observed in satellite imagery; our focus was not on precisely reconstructing every plant of a real environment. While this is arguably important, it would require further analysis (\eg through deep learning) of satellite images and additional data sources, such as coverage maps. To this end we think that procedurally generated vegetation can help to generate training data for more advanced analysis pipelines. Compared to manually placing vegetation, our method provides more control and capabilities for the efficient authoring of vegetation placement for city models.

As an alternative to learning parameters with the neural network pipeline illustrated in Fig.~\ref{fig:networks}, we experimented with learning plant positions with Pix2Pix~\cite{Isola2016ImagetoImageTW} in an end-to-end manner. For this setup we used satellite images as input and images with plant positions and building geometry as an output. The goal was to obtain the plant positions from the images in a post-processing step. Training this network was not successful due to two reasons: for one, it is difficult to obtain ground truth data pairs of satellite images and plant positions. While some datasets contain GPS positions of trees, they only store these positions for trees along streets, which is not useful for learning plant positions of an entire city. Second, the results of the network produced were not satisfactory. We suspect that the ground truth images were to sparse (\ie too few tree positions and building geometry) to provide a meaningful training signal.

A limitation of our current implementation is that we are not able to obtain structural parameters with our learning pipeline. Structural parameters cannot be learned from coverage maps; learning them from top-down satellite images was not successful. Another limitation of our current approach is that we focus on medium and large trees and do not place smaller plants, such as flowers, bushes, or grass. While fixed models of flowers could be placed with our placement strategies (for example by using agent-based models~\cite{Benes:03:TPCG}), there exists no integrated developmental model that would allow us to jointly develop trees and flowers. Therefore, we decided to only simulate the growth response of trees to their environment. Furthermore, we do not model plants that are shaped through advanced topiary. More research would be required to explore how pruning affects growth, \eg for hedges. 

\section{Conclusion and Future Work} \label{sec:future}

We have presented a novel framework for populating synthetic and real urban landscapes with vegetation. To this end we introduced procedural placement models that allow us to realistically generate plant positions and to jointly grow individual plants into individual lots. The key idea to our approach is that complex patterns of vegetation among different zoning types of a city can be factorized into a set of simple placement rules. A PPM implements these rules and -- together with their parameterization -- allows to generate complex patterns of vegetation with high visual fidelity. Moreover, the PPMs are context-sensitive and read the immediate neighborhood which allows us to smooth out abrupt changes in placement.

To populate vegetation into models of real cities we have used a state-of-the-art style-transfer network to translate satellite images to coverage maps of vegetation. These coverage maps allow us to determine the distribution of vegetation within individual lots of a city, which in turn allows us to reconstruct vegetation similar to what can be observed in real data. Instead of reconstructing vegetation at city scale precisely -- which is intractable -- our goal is to generate convincing and plausible details for reconstructing existing cities or for populating entirely new virtual cities with vegetation.

We see a number of avenues for future work. First, it would be interesting to further explore physical functions in an urban context that are affected by vegetation such as heat transfer, shading, wind, and sound barriers. Second, further exploring how neural networks can help to generalize to more diverse urban data and to use them to learn parameters for scene generation seems like a promising direction for future research. Finally, we want to explore enhanced placement strategies to capture more of the variation of vegetation placements that can be observed in real cities.

\bibliographystyle{ACM-Reference-Format}
\bibliography{main}

\appendix
\small

\section{Parameters}
\label{appx:params}

Tab.~\ref{tab:param_values_table} lists the parameter values we used to generate the figures in the paper. 

\begin{table}[h]
\footnotesize
\centering
\vspace{-2mm}
\caption{Parameter values for figures in the paper.}
\vspace{-4mm}
\label{tab:param_values_table}
\scalebox{0.8}{
\begin{tabular}{l|c|ccccccccc|ccccc}
\textbf{Fig.} & Strategy   &  \textbf{$\mu$ }     & \textbf{$\sigma$}    & \textbf{$\tau$} & \textbf{$\beta$} & \textbf{$\kappa$} & \textbf{$\pi$} & \textbf{$\omega$} & \textbf{$\psi$} & \textbf{$\eta$} & \textbf{$\alpha$} & \textbf{$\rho$} & \textbf{$\theta$} &  \textbf{$\lambda$}\\ 
\hline
\ref{fig:positional_params} a      & B & 3.13 & 0.35 & 1.0 & 4.0 & & & & & & & & &\\    
\ref{fig:positional_params} b      & B & 3.13 & 0.35 & 0.8 & 10.0 & & & & & & & & & \\
\ref{fig:positional_params} c      & B & 3.13 & 0.35 & 0.4 & 10.0 & & & & & & & & &\\
\ref{fig:positional_params} d      & C & 3.13 & 0.35 & 1.0 & & 10.0 & 1 & & & & & & &\\
\ref{fig:positional_params} e      & C & 3.13 & 0.35 & 1.0 & & 10.0 & 3 & & & & & & &\\
\ref{fig:positional_params} f      & C & 3.13 & 0.35 & 1.0 & & 15.0 & 3 & & & & & & &\\
\ref{fig:positional_params} g      & I & 3.13 & 0.00 & 1.0 & & & & 9.0 & 0.00 & 30\textdegree & & & &\\
\ref{fig:positional_params} h      & I & 3.13 & 0.00 & 1.0 & & & & 9.0 & 0.30 & 30\textdegree & & & &\\
\ref{fig:positional_params} i      & I & 3.13 & 0.00 & 1.0 & & & & 9.0 & 0.55 & 30\textdegree & & & &\\
\ref{fig:positional_params} j      & R & 3.13 & 0.35 & 0.2 & & & & & & & & & &\\
\ref{fig:positional_params} k      & R & 3.13 & 0.35 & 0.4 & & & & & & & & & &\\
\ref{fig:positional_params} l      & R & 3.13 & 0.35 & 1.0 & & & & & & & & & &\\
\ref{fig:structural} a             & R & 3.4 & 0.15 & 0.2 & & & & & & & 14 & 0.0 & 0.0 & 4\\
\ref{fig:structural} b             & R & 3.4 & 0.15 & 0.2  & & & & & & & 20 & 0.0 & 0.0 & 4\\
\ref{fig:structural} c             & R & 3.4 & 0.15 & 0.2 & & & & & & & 30 & 0.0 & 0.0 & 4\\
\ref{fig:structural} d             & B & 2.2 & 0.0 & 0.9 & 10 & & & & & & 20 & 1.0 & 0.0 & 4\\
\ref{fig:structural} e             & B & 2.2 & 0.0 & 0.9 & 10 & & & & & & 20 & 0.7 & 0.0 & \\
\ref{fig:structural} f             & B & 2.2 & 0.0 & 0.9 & 10 & & & & & & 20 & 0.5 & 0.0 & 4\\
\ref{fig:structural} g             & C & 3.2 & 0.25 & 1.0 & & 10.0 & 3 & & & & 16 & 0.0 & 0.0 & 4\\
\ref{fig:structural} h             & C & 3.2 & 0.25 & 1.0 & & 10.0 & 3 & & & & 16 & 0.1 & 0.3 & 4\\
\ref{fig:structural} i             & C & 3.2 & 0.25 & 1.0 & & 10.0 & 3 & & & & 16 & 0.2 & 0.4 & 4\\
\ref{fig:growth_of_building}       & C & 3.5 & 0.35  & 1.0 & & 20.0 & 18 &  & & & 16 & 0.0 & 0.0 & 1\\
\hline    
\end{tabular}
}
\vspace{-4mm}
\end{table} 

\section{Satellite and Coverage Map Data}
\label{apx:map_images}
\begin{figure}[h]
    \centering
    \includegraphics[width=\linewidth]{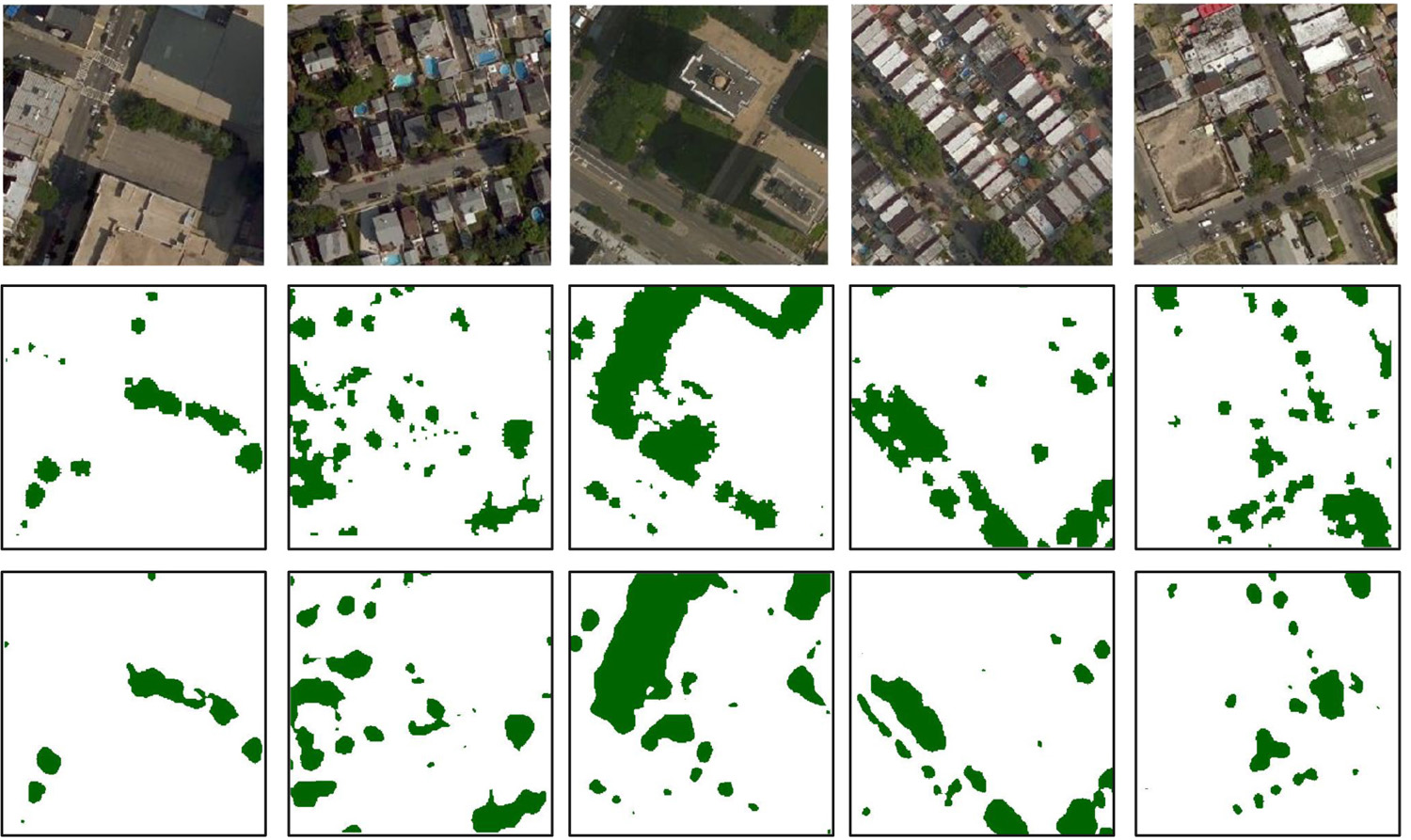}\\
    \vspace{-3mm}
    \caption{Learning of coverage maps: we use satellite images (top) and ground truth coverage maps (middle) from NYC Open Data to train a neural network for style-transfer. After training the network is able to predict coverage maps (bottom) from satellite images. }
    \label{fig:semantic_maps}
\end{figure}

Examples of satellite images (top), ground truth coverage maps (middle) and predicted coverage maps (bottom) are shown in Fig.~\ref{fig:semantic_maps}.

\end{document}